\documentclass[11pt, a4paper]{article}
\usepackage{aasmacros}
\usepackage{jcappub}

\usepackage{booktabs}
\usepackage{multirow}
\usepackage{amssymb}
\usepackage[export]{adjustbox}
\usepackage{floatrow}
\newfloatcommand{capbtabbox}{table}[][3.5in]
\newfloatcommand{capbfigbox}{figure}[][2.3in]
\usepackage{blindtext}
\usepackage{amsmath}
\usepackage[normalem]{ulem}

\newcommand{\Fermi}{\textit{Fermi}}

\newcommand{\cf}{cf.~}

\newcommand{\Msun}{{\rm M}_\odot}

\begin{document}

\title{Simulated Milky Way analogues: implications for dark matter indirect searches}

\author[a]{Francesca Calore,}
\emailAdd{f.calore@uva.nl}
\author[a]{Nassim Bozorgnia,}
\author[a,b]{Mark Lovell,}
\author[a]{Gianfranco Bertone,}
\author[c]{Matthieu Schaller,}
\author[c]{Carlos S. Frenk,}
\author[d]{Robert A. Crain,}
\author[e]{Joop Schaye,}
\author[c]{Tom Theuns}
\author[c]{\& James W. Trayford}

\affiliation[a]{GRAPPA, University of Amsterdam, Science Park 904,
  1090 GL Amsterdam, Netherlands} 
\affiliation[b]{Instituut-Lorentz
  for Theoretical Physics, Niels Bohrweg 2, NL-2333 CA Leiden,
  Netherlands}
\affiliation[c]{Institute for Computational Cosmology, Durham
  University, South Road, Durham DH1 3LE, UK}
\affiliation[d]{Astrophysics Research Institute, Liverpool John Moores University, 146 Brownlow Hill, Liverpool L3 5RF, UK}
\affiliation[e]{Leiden Observatory, Leiden University, PO Box 9513, 
  NL-2300 RA Leiden, Netherlands}

\abstract{ We study high-resolution hydrodynamic simulations of Milky Way type
  galaxies obtained within the ``Evolution and Assembly of GaLaxies and their
  Environments'' (EAGLE) project, and identify those that best
  satisfy observational constraints on the Milky Way total stellar mass, rotation
  curve, and galaxy shape.  Contrary to mock galaxies selected on the basis of their
  total virial mass, the Milky Way analogues so identified consistently exhibit very
  similar dark matter profiles inside the solar circle, therefore enabling more accurate
  predictions for indirect dark matter searches. We find in particular
  that high resolution simulated haloes satisfying observational constraints exhibit, 
  within the inner few kiloparsecs, dark
  matter profiles shallower than those required  
  to explain the so-called {\it Fermi~GeV excess} via dark matter
  annihilation.
  }

\maketitle


\section{Introduction}
\label{sec:intro}
Discovering the nature of dark matter (DM) is one of the main challenges for physics
today.  We have evidence of the DM gravitational interactions at different scales,
from galactic up to cosmological scales (see
e.g.~\cite{Bertone:2010zza,Jungman:1995df,Bergstrom00,Bertone:2004pz}), but its
nature remains a mystery.  Weakly Interacting Massive Particles (WIMPs) are among the
most well-motivated DM particle candidates, and they are currently searched for with
three detection strategies: (a) {\it direct} detection, based on the measurement of
the recoil energy of nuclei hit by DM particles in underground
experiments, (b) {\it indirect detection}, through the search for secondary particles
produced in the annihilation or decay of DM, and (c) the search for new particles 
beyond the standard model of particle physics at accelerators, in particular at the Large Hadron Collider.

\medskip 

The prospects for discovering DM particles with direct and indirect searches strongly
depend on the distribution of DM in the Milky Way (MW).  Recently an extensive compilation
of measurements of the Milky Way rotation curve -- i.e. of the circular velocity of astrophysical
tracers as a function of the distance from the Galactic centre (GC) -- has confirmed the
existence of large amounts of DM within the solar
circle~\cite{Iocco:2015xga}, and enabled researchers to estimate the DM spatial density profile with
parametric~\cite{2015arXiv150406324P} and non-parametric
methods~\cite{2015ApJ...803L...3P,2007MNRAS.378...41S}.

The DM density profile, $\rho(R)$ (where $R$ is the distance from the GC), is
a crucial ingredient for predicting the intensity flux and anisotropy of gamma rays produced by the
annihilation of WIMPs in the halo of the MW, since the annihilation rate depends on
the square of the DM particle number density.  While the DM density profile at
galactocentric distances $R > 5 - 6$ kpc is relatively well constrained by the
analysis of kinematical data of the MW rotation curve, the DM profile in the inner
few kiloparsecs from the GC is subject to large
uncertainties~\cite{2015ApJ...803L...3P}.  In the absence of observational constraints,
the most profitable recourse is examination of N-body, and, more recently, hydrodynamic
simulations. Pure DM N-body simulations predict a DM density profile that behaves
like $r^{-\gamma}$, where $\gamma \approx 1$ in the few inner kiloparsecs, as encoded in
the so-called Navarro-Frenk-White (NFW) profile~\cite{Navarro:1996gj}. 
Baryonic processes can either lower or increase the DM density in the 
inner halo~\cite{Navarro:1996bv,Weinberg:2001gm,Mashchenko:2006dm,
Pontzen:2011ty,Mollitor:2014ara,Blumenthal:1985qy,Gnedin:2004cx}.
However, the size of those effects is still a matter of debate.

\medskip 

In a recent exciting development for indirect dark matter searches, 
an unexplained excess of gamma rays collected with the Large Area Telescope (LAT) 
-- aboard the \Fermi~satellite -- from the
centre of our Galaxy has been discovered over and above the standard adopted astrophysical
background~\cite{Goodenough:2009gk, Hooper:2010mq, Hooper:2011ti, Abazajian:2012pn,
  Gordon:2013vta, Hooper:2013rwa, Abazajian:2014fta, Daylan:2014rsa, Calore:2014xka,
  fermigc}.  In particular, ref.~\cite{Calore:2014xka} re-assessed the spectral and
morphological properties of the GeV excess, taking into account background model
systematics associated with the Galactic diffuse emission modelling.  The striking similarity
of the observed gamma-ray excess with the signal predicted from the annihilation of
DM particles in the halo of the MW makes the DM interpretation very appealing (see
e.g. ~\cite{Calore:2014nla,Daylan:2014rsa, Abazajian:2014fta,Caron:2015wda}),
although other viable astrophysical explanations have been put
forward~\cite{Hooper:2013nhl, Cholis:2014lta, Petrovic:2014xra,Yuan:2014rca,
  O'Leary:2015gfa, Carlson:2014cwa, Petrovic:2014uda,Cholis:2015dea,Gaggero:2015nsa}.
  
Among the alternative GeV excess sources, the possibility that the excess originates from a series of leptonic
outbursts which occurred $\sim 0.1 - 1$ yr ago has recently been demonstrated to be a
 viable scenario but a quite unlikely one~\cite{Cholis:2015dea} given
the set of parameters needed to fully account for the spectral \emph{and}
morphological properties of the GeV excess emission.  On the other hand, the vigorously
debated interpretation in terms of the unresolved emission from very dim
sources, such as for example pulsars~\cite{O'Leary:2015gfa} and milli-second
pulsars~\cite{Hooper:2013nhl, Cholis:2014lta, Petrovic:2014xra,Yuan:2014rca}, has
been very recently corroborated by two independent works~\cite{Lee:2015fea,Bartels:2015aea}.  
While it is clear that the disc population of unresolved pulsars and milli-second pulsars
cannot contribute more than 10\% to the excess emission~\cite{Calore:2014oga}, the
contribution from a new population associated with the Galactic bulge seems instead
to be sufficient to explain the full signal~\cite{Lee:2015fea,Bartels:2015aea}.

Given the lively debate on possible explanations, and the difficulty of firmly
confirming or refuting them, it is crucial to fully exploit state of the art simulations
to examine what is the expected DM profile of MW-like galaxies, in order to 
refine predictions for the DM annihilation gamma-ray flux. 
We will therefore compare the predictions from the set of selected MW-like
galaxies with the GeV excess gamma-ray measurements.

\medskip

In this work, we study the distribution of DM in MW-like galaxies
simulated within the EAGLE project \cite{Schaye:2015,Crain:2015} -- a suite of
cosmological, hydrodynamic simulations calibrated to reproduce the observed
distribution of stellar masses and sizes of
low-redshift galaxies and designed to address many outstanding issues
in galaxy formation such as metallicities of galaxies, properties of the
intergalactic medium and the effect of feedback on scales ranging from dwarf
galaxies up to giant ellipticals.  We consider at first all galaxies within haloes
with virial mass in the range $\mathcal{O}(10^{12} - 10^{14}) \, \Msun$, and we
post-process this sample of MW analogues by requiring that they satisfy
observational constraints on the Galactic rotation curve~\cite{Iocco:2015xga}, 
the total stellar mass and the presence of a dominant disc
in the stellar component.

We then evaluate the DM density profile of the final set of what we define to be
\emph{MW-like galaxies} and discuss the prospects for DM indirect
detection (the implications for direct detection will be discussed in an upcoming paper~\cite{Calore:XXX2}).  In particular, we discuss the
implications of the resulting DM density profiles for the DM interpretation of the \Fermi~GeV excess.

\bigskip 

After presenting the set of cosmological hydrodynamic simulations used in
section~\ref{sec:simulation}, we will describe our selection procedure and derive the
set of MW-like galaxies in our samples in
section~\ref{sec:selection}.  We will dedicate section~\ref{sec:rhodm} to the
analysis of the DM density profiles of the set of MW-like galaxies and finally
section~\ref{sec:idm} to the discussion of the implications for the GeV excess
emission.  We conclude in section~\ref{sec:conclusion}.  Appendices \ref{app:M200}
and \ref{app:v0R0} contain additional material supporting our findings.

\section{Simulations}
\label{sec:simulation}
In this section we briefly describe the set of simulations used,
which form part of the EAGLE project \citep{Schaye:2015,Crain:2015}.  The EAGLE
simulations were performed using a modified version of the \textsc{P-gadget3}
Tree-Smoothed Particle Hydrodynamics (SPH) code~\cite{Springel:2008b} that
  has been modified to use the state-of-the-art SPH flavour
  of~\cite{Hopkins:2013} (whose impact on the galaxy population is
  discussed by \cite{Schaller:2015sph}). 
  The cosmological parameters were chosen to be those derived from
  the analysis of the \emph{Planck} 2013 measurements~\cite{Planck:2014} 
  and they have the following values: $\Omega_{m}=0.307$,
$\Omega_{\Lambda}=0.693$, $\Omega_{b}=0.0482$, $h=0.678$, $\sigma_{8}=0.83$, and
$n_{s}=0.961$. The simulations were run at two different resolutions, which we refer
to as intermediate (EAGLE IR) and high resolution (EAGLE HR).  The former was run in
a series of cosmological volumes up to a maximum of 100 Mpc on one side, and the
latter up to 25 Mpc.  The simulations start at z=127 with an equal number of gas and
DM particles whose masses are given in table~\ref{Tab:sims}.  Gas particles have a finite
probability to be turned into star particles that increases with the gas pressure, 
such that the local star formation law \cite{Kennicutt:1998} is
  reproduced. Each star particle represents a simple stellar population, and
inherits the mass and element abundances of the parent gas
particle. The properties of the two EAGLE runs used in this study (EAGLE IR and EAGLE HR)
are reproduced from~\cite{Schaye:2015} in table~\ref{Tab:sims}.  In addition to the
EAGLE cosmological volumes, we also make use of simulations of the APOSTLE
project~\cite{2014arXiv1412.2748S,2015arXiv150703643F}.  These simulations use the
same code as the EAGLE project applied to zoomed regions containing a close
pair of $\sim10^{12}\Msun$ DM haloes that could host our MW galaxy and M31, i.e. be
an analogue of the Local Group.  We use twelve APOSTLE volumes simulated at similar
resolution to EAGLE HR, which as a group we denote APOSTLE IR (for ``intermediate"
resolution). In addition, APOSTLE consists of other two re-simulations at ten times higher mass
resolution, denoted generically as APOSTLE HR.  Their simulation details are also
included in table~\ref{Tab:sims}.  One minor difference from the EAGLE cosmological
volumes is that they use the WMAP-7 cosmological parameters: $\Omega_{m}=0.272$,
$\Omega_{\Lambda}=0.728$, $\Omega_{b}=0.0455$, $h=0.704$, $\sigma_{8}=0.81$, and
$n_{s}=0.967$.

\begin{table}
    \centering
    \begin{tabular}{|c|c|c|c|c|c|}
      \hline
       Name & L (Mpc) & N & $m_{\rm g}~(\Msun)$ & $m_{\rm dm}~(\Msun)$ & $\epsilon$ (pc) \\
      \hline
           EAGLE HR & 25 &   $2\times752^{3}$ & $2.26\times10^{5}$ & $1.21\times10^{6}$ & 350 \\
           EAGLE IR & 100 & $2\times1504^{3}$ & $1.81\times10^{6}$ & $9.70\times10^{6}$ & 700 \\
           APOSTLE IR    & -- & -- & $1.3\times10^{5}$ & $5.9\times10^{5}$ & 308 \\          
           APOSTLE HR (I) & -- & -- & $1.0\times10^{4}$ & $5.0\times10^{4}$ & 134 \\
           APOSTLE HR (II)& -- & -- & $5.0\times10^{3}$ & $2.5\times10^{4}$ & 134 \\ 
      \hline
    \end{tabular}
    \caption{Parameters of the simulations discussed in this paper. $L$ is the
      comoving sidelength of the simulation cube, $N$ the number of simulation
      particles prior to splitting, $m_{\rm g}$ the initial gas particle mass,
      $m_{\rm dm}$ the DM particle mass, and $\epsilon$ the Plummer-equivalent
      physical softening length. The resolution limit is usually taken to be
      2.8$\times \epsilon$, i.e.~1.96, 0.98 and 0.87 kpc for EAGLE IR, EAGLE HR and
      APOSTLE IR, respectively.}
    \label{Tab:sims}
  \end{table}

\medskip

To select MW-like galaxy candidates, we first extract all galaxies
located at the minimum of their halo potential wells as returned by the
  \textsc{Subfind} algorithm \cite{Springel:2001} (i.e. we exclude satellite galaxies) in haloes of
virial mass, $M_{200}$, in the range $5\times10^{11}<M_{200}/\Msun<1\times10^{14}$,
where $M_{200}$ is defined as the mass enclosed within the sphere that contains a
mean density 200 times the critical density. We extend our range to the most massive
haloes because the model underpredicts slightly the stellar masses contained within
haloes of $M_{200}\sim10^{12}\Msun$~\cite{Schaye:2015}, and therefore MW mass
galaxies are found in haloes that are slightly more massive than is inferred from
abundance matching results, e.g.~\cite{2013ApJ...770...57B}.  We show in
appendix~\ref{app:M200} that this mismatch between halo mass and stellar mass has
little effect on our results.\footnote{Appendices refer to the EAGLE set of
  simulations. However, the conclusions reached there can be safely extended to the
  APOSTLE IR run.}

\section{Selection of Milky Way-like galaxies}
\label{sec:selection}
We consider the simulation runs EAGLE IR, EAGLE HR and APOSTLE IR described in section~\ref{sec:simulation}.
We start from the corresponding subsets of galaxies at the centre of haloes with $5\times10^{11}<M_{200}/\Msun<1\times10^{14}$.
The initial sets are composed of 2411, 61 and 24 objects for the EAGLE IR, EAGLE HR and APOSTLE IR run, respectively.

\medskip 

In this section we aim to select the galaxies that most closely resemble the MW.
Our definition of MW-like galaxies is based on a minimal set of criteria that the simulated haloes
should satisfy.  In particular, for a simulated halo to host a good MW-like galaxy, we require that:
\begin{itemize}

\item[(i)] The simulated rotation curve fits well the observed MW kinematical data
  in ref.~\cite{Iocco:2015xga}.  We explain the method followed to derive the rotation
  curves from the simulation, the data used in the analysis and the goodness of fit
  definition in section~\ref{sec:rotcurves}. 

\item[(ii)] The total stellar mass of the simulated galaxies is within the 3$\sigma$
  MW range derived from observations,
  $4.5 \times10^{10}<M_{*}/\Msun<8.3 \times10^{10}$~\cite{McMillan:2011wd}: 335, 12, and 2
  galaxies satisfy this constraint in the EAGLE IR, EAGLE HR and APOSTLE IR run
  respectively.\footnote{We note that for APOSTLE IR one of the two galaxies
   falls on the lower boundary of the 3$\sigma$ observed range.
    In order to have more than one galaxy in the final selection, we keep this one as well.
}

\item[(iii)] The galaxies contain a substantial stellar disc component. See section~\ref{sec:disc}.

\end{itemize}

The three criteria listed above define our MW-like galaxies. The final sets of
\emph{good} objects are presented in section~\ref{sec:selection} for the EAGLE IR, EAGLE HR
and APOSTLE IR simulations.

We are aware of the fact that besides structural parameters, such as the rotation
curve used in the present work, more criteria should be imposed to identify truly
MW-like galaxies, such as brightness profile constraints, star formation history,
metallicity gradient, and disc scale length/height. However, we stress that our main
purpose is to derive implications for indirect DM searches rather than to test the
ability of the EAGLE simulation code to reproduce the MW.  
Indirect detection prospects depend on the shape of
the DM profile, which turns out to be almost universal for the simulated objects in
EAGLE IR, EAGLE HR and APOSTLE IR as shown in ref.~\cite{Schaller:2014uwa} and more thoroughly in
section~\ref{sec:rhodm}.
The method presented here is general and can be applied to future simulations.

\subsection{Observed Milky Way rotation curve and goodness of fit}
\label{sec:rotcurves}
Recently, ref.~\cite{Iocco:2015xga} collated a large amount of
observational measurements of the MW rotation curve and compared
these with the expectations from a large set of baryonic models,
finding robust evidence for DM even within the inner 5 -- 6 kpc of the Galaxy.  We
make use of the recent compilation of kinematic tracers presented
in~\cite{Iocco:2015xga}.  We adopt a local circular velocity of $v_0 = 230$ km/s, a
local galactocentric distance of $R_0 = 8$ kpc, and the component of the solar
peculiar velocity in the direction of the Galactic rotation, $V_{\odot} = 12.24$ km/s
\cite{Schoenrich:2010}, as fiducial values for the analysis presented below.
However, owing to the thorough discussion by~\cite{Iocco:2015xga} about
the uncertainty in the rotation curve data due to $v_0$ and $R_0$, we dedicate
appendix~\ref{app:v0R0} to the scrutiny of possible
variations in the results due to different choices of $v_0$ and $R_0$ and we show
that our main conclusions remain unchanged.

The observational data are provided as constraints on the angular circular velocity,
$\omega_c (R)=v_c(R)/R$, and the galactocentric distance $R$. Here $v_c(R)$ is the
circular velocity at distance $R$. As explained by~\cite{Iocco:2015xga}, using the
angular circular velocity for fitting purposes is more convenient than working with
$v_c(R)$, since the errors of $\omega_c(R)$ and $R$ are not correlated, while this is
not the case for $v_c(R)$ and $R$.

\medskip 

\begin{figure}[t!]
    \begin{center}
        \includegraphics[width=0.48\linewidth]{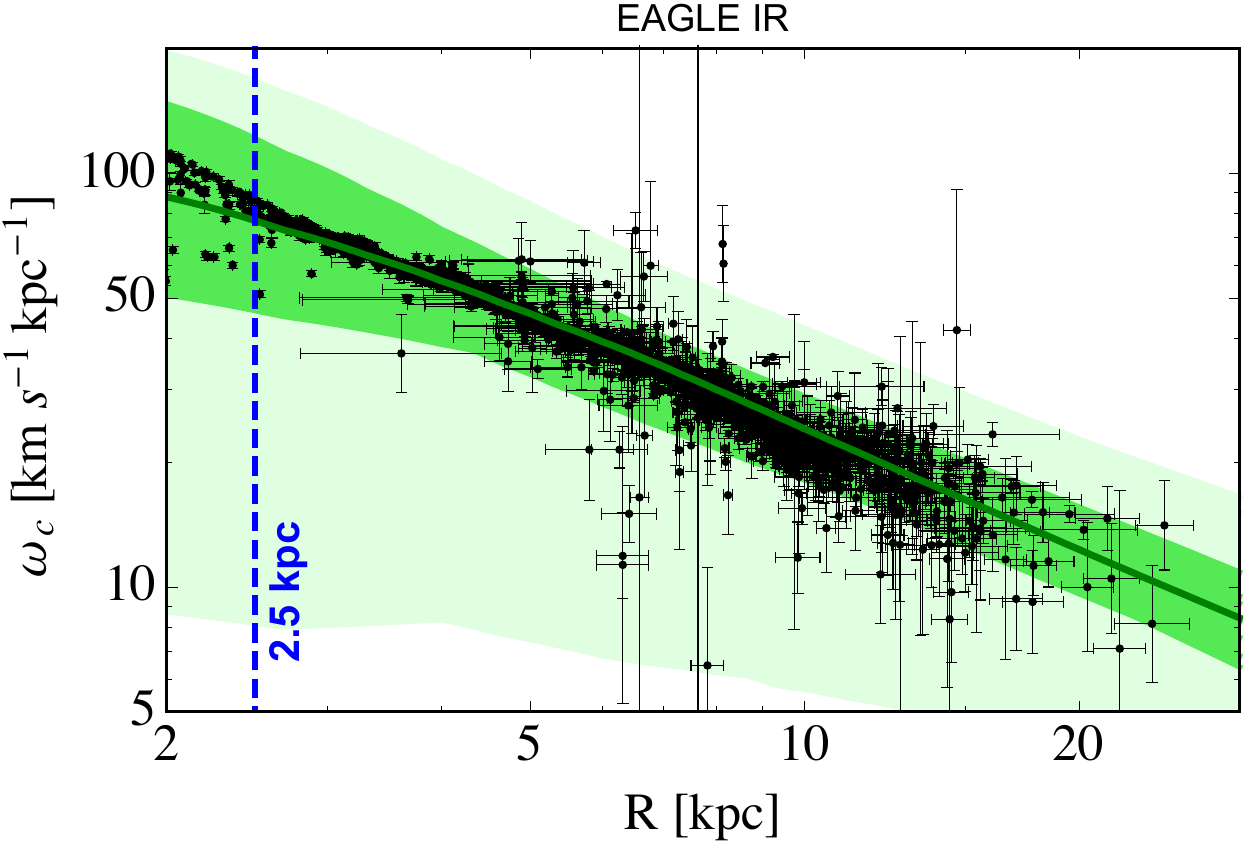}
        \includegraphics[width=0.48\linewidth]{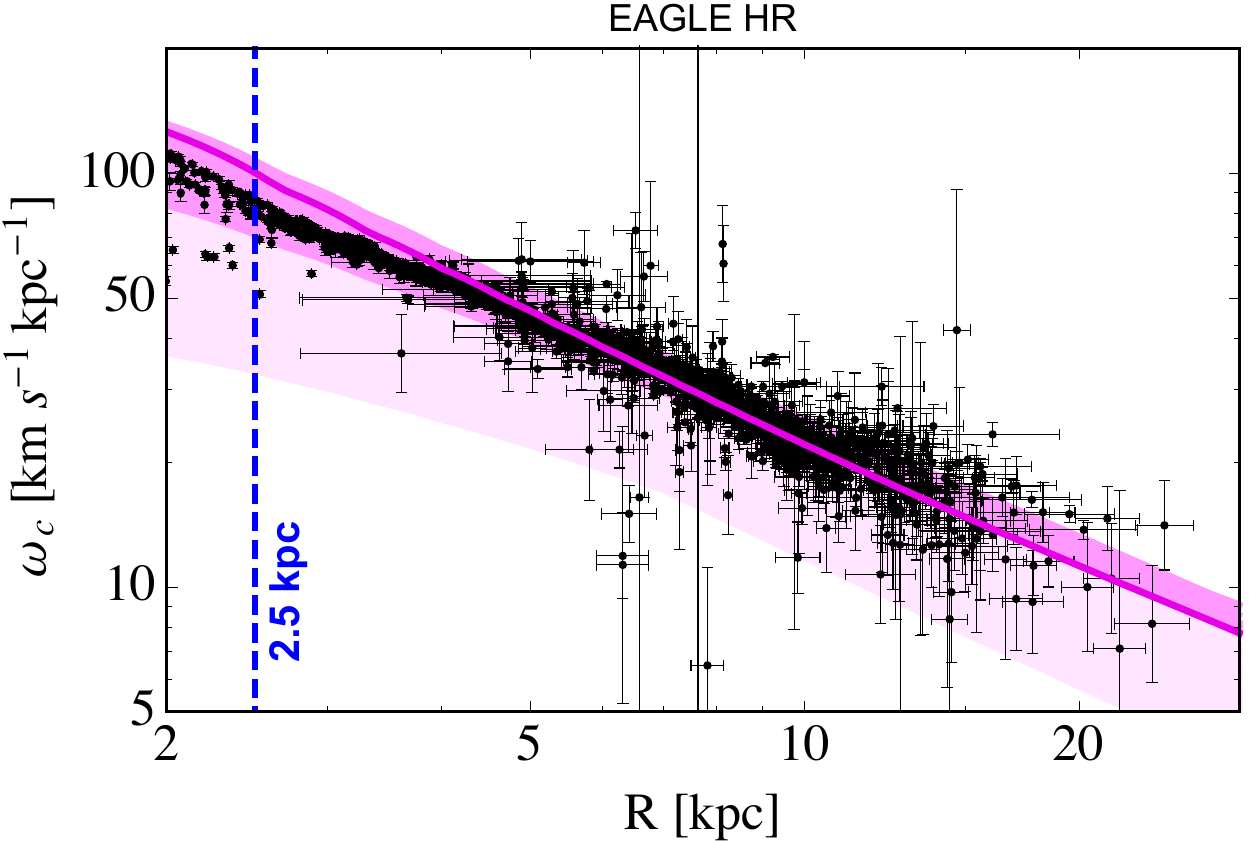}\\
        \includegraphics[width=0.48\linewidth]{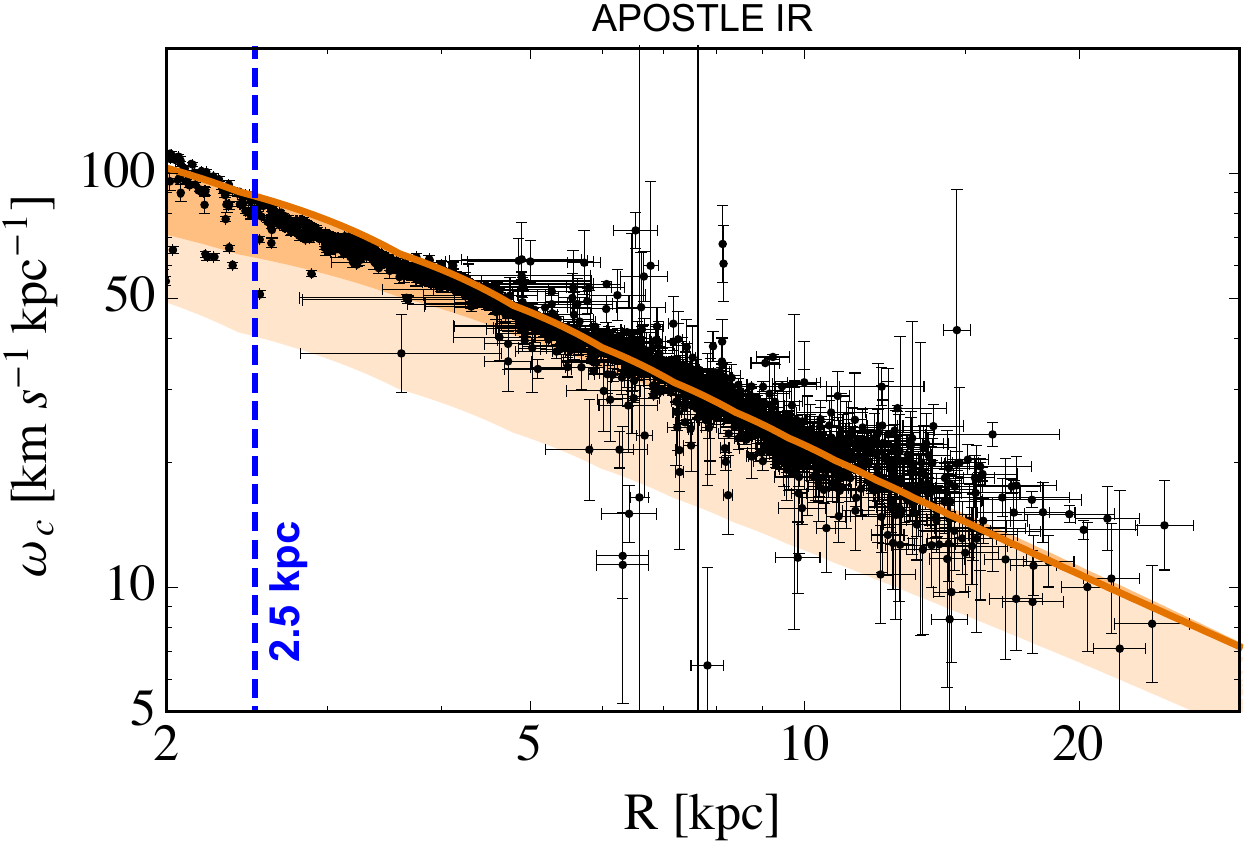}
    \end{center}
    \caption{Kinematical data from~\cite{Iocco:2015xga} used in this work
      (\emph{black} points and error bars) and rotation curves of simulated haloes
      for the  EAGLE IR (\emph{top left} panel),  EAGLE HR (\emph{top right} panel),
       and APOSTLE IR (\emph{bottom centre} panel) runs. The
      \emph{green}, \emph{magenta}, and \emph{orange} curves correspond to galaxies which fit our
      criteria as defined in section~\ref{sec:selection} and give the best fit
      (lowest $\chi^2$ in eq.~\eqref{eq:chi2vc}) in the EAGLE IR, EAGLE HR, and APOSTLE IR runs, respectively.
      The \emph{light} and \emph{dark} coloured bands correspond to all
      simulated objects ($5\times10^{11}<M_{200}/\Msun<1\times10^{14}$)
      and those which, in addition, satisfy the observed MW stellar mass range
      ($4.54\times10^{10}<M_{*}/\Msun<8.32\times10^{10}$), respectively.  The
      vertical \emph{dashed blue} line marks the minimal radius considered in the
      fitting procedure (see text for further details).}
    \label{fig:rotcurves}
\end{figure}

The circular velocity, $v_{c}(R),$ is defined as the velocity of a test particle on a
circular orbit at radius $R$ from the GC. By equating the centripetal and
gravitational forces on the test particle, it is simple to show that, for a spherically symmetric matter distribution:

\begin{equation}
v_c(R) = \sqrt{\frac{G M(< R)}{R}} \, ,
\label{eq:vc}
\end{equation}
where $M(<R)$ is the total mass enclosed within radius $R$, and $G$ is the universal
gravitational constant.  In figure~\ref{fig:rotcurves}, we display the kinematical
data in the plane ($\omega_c$, $R$) together with the rotation curves predicted by
\emph{all} haloes in the EAGLE IR (top left panel), EAGLE HR (top right panel), and APOSTLE
IR (bottom panel) simulation runs, satisfying our halo mass constraint.  From
figure~\ref{fig:rotcurves}, it is evident that there is a wide range of variation in
the rotation curves from the simulation when considering all objects whose halo mass,
$M_{200}$, lies in the selected MW mass range,
$5\times10^{11}<M_{200}/\Msun<1\times10^{14}$. However, by forcing the haloes to have
the correct total stellar mass (in the 3$\sigma$ observed range for the MW), the
number of good objects is reduced significantly: from 2411 to 335 for  EAGLE IR, from 61
to 12 for EAGLE HR, and from 24 to 2 for  APOSTLE IR. We also
display the haloes that give the best fit to the kinematical data (see below for
further details).  Therefore, classifying a halo with the correct halo mass as
MW-like is too simplistic a criterion, and thus will often fail to reproduce the MW
kinematical data. For this reason, we extend the definition of MW-like galaxies to
account for the agreement with the observed Galactic rotation curve.\footnote{More realistically 
the requirement of spherical symmetry is not guaranteed to be a good approximation, and 
thus Equation~\ref{eq:vc} may not be appropriate. As a check, for the APOSTLE IR galaxies 
we calculated a series of rotation curves by summing
 up the gravitational forces due to all of the simulation particles in the box. We found
  that the difference in rotation curve amplitude between the full calculation and that 
  obtained from the spherical symmetry is typically less
   than 10\% at all radii. Moreover, we checked that the $\chi^2$ values of these new
    rotation curves (for the fit to the  measured MW rotation curve) are not significantly
     different from the ones obtained by assuming spherical symmetry. Most importantly, 
     the ordering of haloes based on $\chi^2$ values stays the same. We therefore 
     conclude that our assumption of spherical symmetry is appropriate for this study. }

\medskip
 
To derive the goodness of fit of the simulated rotation curve to the observed data, it is convenient to 
work with the reduced quantities $x  \equiv R/R_0$ and  $y \equiv \omega_c/\omega_0 -1$~\cite{Iocco:2015xga}, 
where $\omega_0 \equiv v_0/R_0$.
The two-variable $\chi^2$ can be written as:
\begin{equation}
\chi^2=\sum_i  \frac{ (y_i -  \hat{y}(x_i))^2 }{ \sigma_{y_i}^2 +  (dy_i/dx_i)^2 \, \sigma_{x_i}^2 } \, , 
\label{eq:chi2vc}
\end{equation}
where $i$ runs over the observational data points considered, and $\hat{y}(x_i)$ is
the simulated rotation curve evaluated at $x_i = R_i/R_0$. Both experimental errors
in $x$ ($\sigma_{x_i}$) and $y$ ($\sigma_{y_i}$) are considered and obtained through
standard error propagation from the errors in $\omega_c$ and $R$.  For the fit, we
consider measurements at $R>2.5$ kpc. Indeed, the data derived
by~\cite{Iocco:2015xga} assume circular orbits for the tracers and this approximation
can break at small radii due to the effect of the Galactic bulge gravitational
potential. Note also that $2.5$ kpc is larger than the gravitational
  softening length used in our simulations.

\bigskip

\begin{figure}
    \begin{center}
        \includegraphics[width=0.48\linewidth]{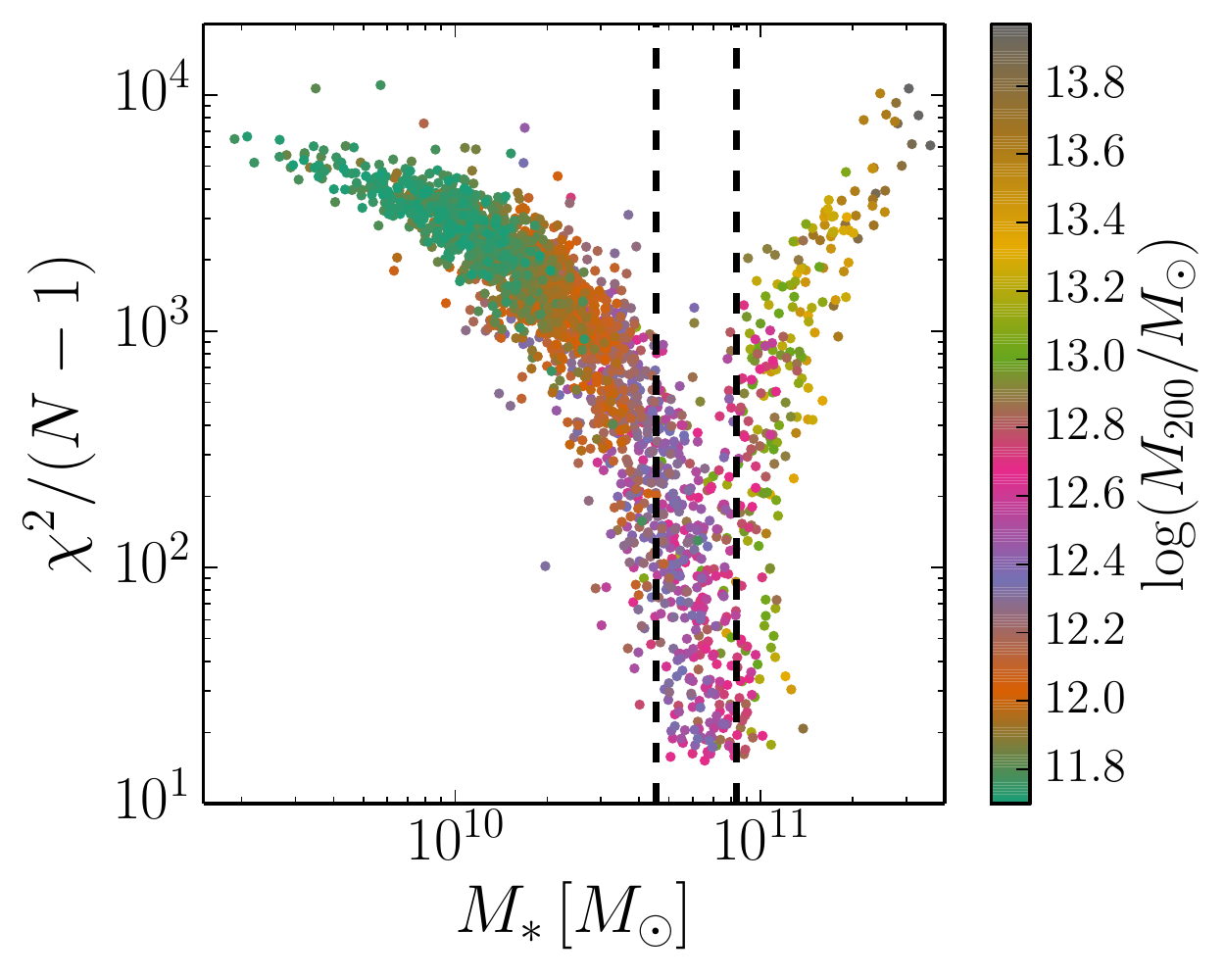} 
        \includegraphics[width=0.48\linewidth]{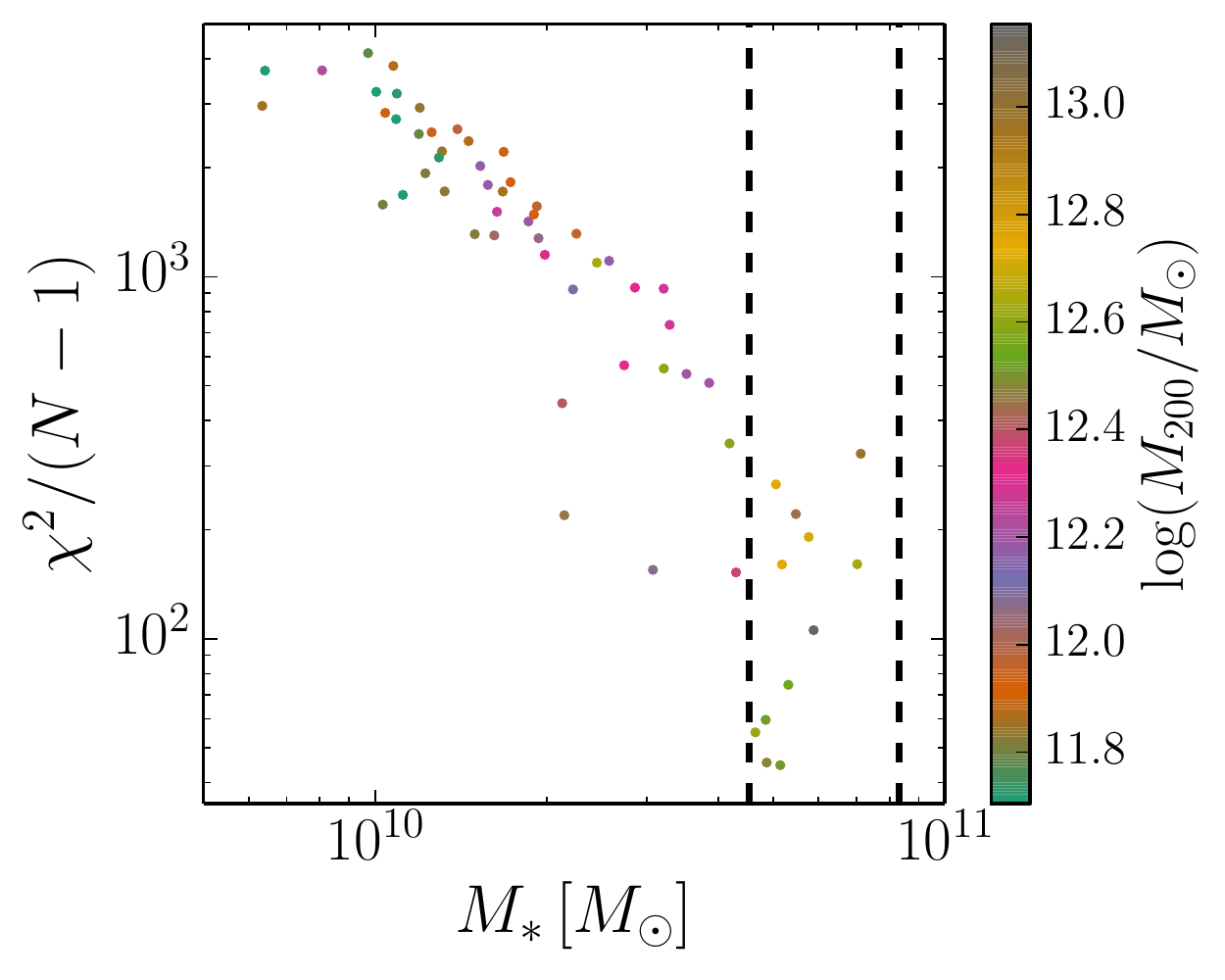} \\
        \includegraphics[width=0.48\linewidth]{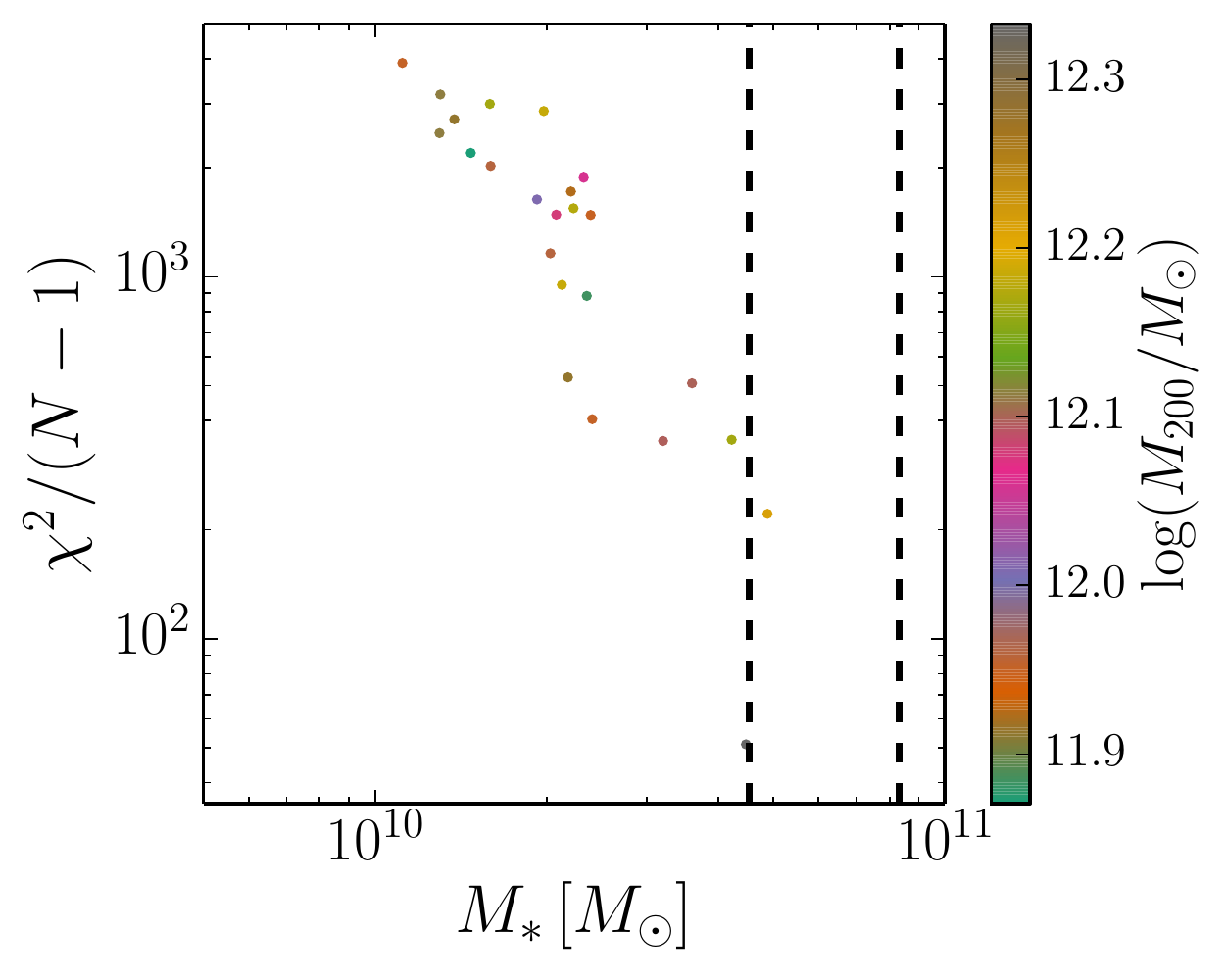} 
    \end{center}
    \caption{Reduced $\chi^2$, $\chi^2/(N-1)$, versus total stellar mass for EAGLE IR
      (\emph{top left} panel), EAGLE HR (\emph{top right} panel) and APOSTLE IR 
      (\emph{bottom} panel). The \emph{coloured} dots
      indicate the whole set of 2411, 61 and 24 simulated haloes in EAGLE IR, EAGLE HR and
      APOSTLE IR; of
      those, 335, 12 and 2 galaxies lie in the 3$\sigma$ measured range (\emph{black
        dashed} vertical lines) of the MW total stellar mass~\cite{McMillan:2011wd}.
      The colour-bar shows the distribution of halo mass, $M_{200}$.
      }
    \label{fig:chi2Mstar}
\end{figure}

In figure~\ref{fig:chi2Mstar}, we show the reduced $\chi^2$ versus the total stellar
mass, with N=2687 the total number of observational data points for the three sets of
simulations.  From the top left panel, which refers to EAGLE IR run, it is evident
that the global minimum of the $\chi^2/(N-1)$ distribution naturally occurs
within the 3$\sigma$ measured range of the MW total stellar
mass~\cite{McMillan:2011wd}. In other words, a good match of the simulation with the
measured MW rotation curve is given by galaxies that have the correct MW stellar
mass.  In general, the contribution to rotation curves from stars (which largely
dominate over the gas in the total baryonic component) is larger than the DM
contribution up to $R \lesssim$ 5 kpc.  We note that by performing the analysis with
a distance cut at 5 kpc our results remain unchanged.
 
The halo masses of the simulated galaxies with correct stellar mass and in the minimum reduced
$\chi^2$ region are larger than expected from observations of the MW, as for example
$M_{200,\rm MW} = 1.2^{+0.7}_ {-0.4} \times 10^{12} \,
\Msun$~\cite{Busha:2010sg}. This result might indicate that the feedback in EAGLE
simulated haloes in this mass range is slightly too efficient, and thus the stellar
mass per unit halo is suppressed with respect to 
that inferred from estimates of the MW stellar and halo masses. This shortcoming is reflected in the stellar luminosity function,
 in that EAGLE underpredicts the abundance of galaxies with stellar masses $\sim 2-8 \times 10^{10} M_{\odot}$ relative to what is observed in galaxy
 surveys \cite{Schaye:2015}.  We also note that the total MW halo mass
is affected by large uncertainties, with estimates based on kinematics of satellites, abundance matching,
and the local Hubble flow, yielding somewhat discrepant results
\cite{Peebles:2011,Behroozi:2013,Forero:2013,Barber:2014}. However, in
appendix~\ref{app:M200}, we show that the quantities relevant for DM indirect
detection, and, in particular, for the implications on the GeV excess interpretation,
are not affected by the large halo mass. We are thus confident that this
mismatch between halo mass and stellar mass has little effect on our final results.

\subsection{Morphology of simulated galaxies: disc and spheroid}
\label{sec:disc}
\begin{figure}[t!]
    \begin{center}
        \includegraphics[width=0.48\linewidth]{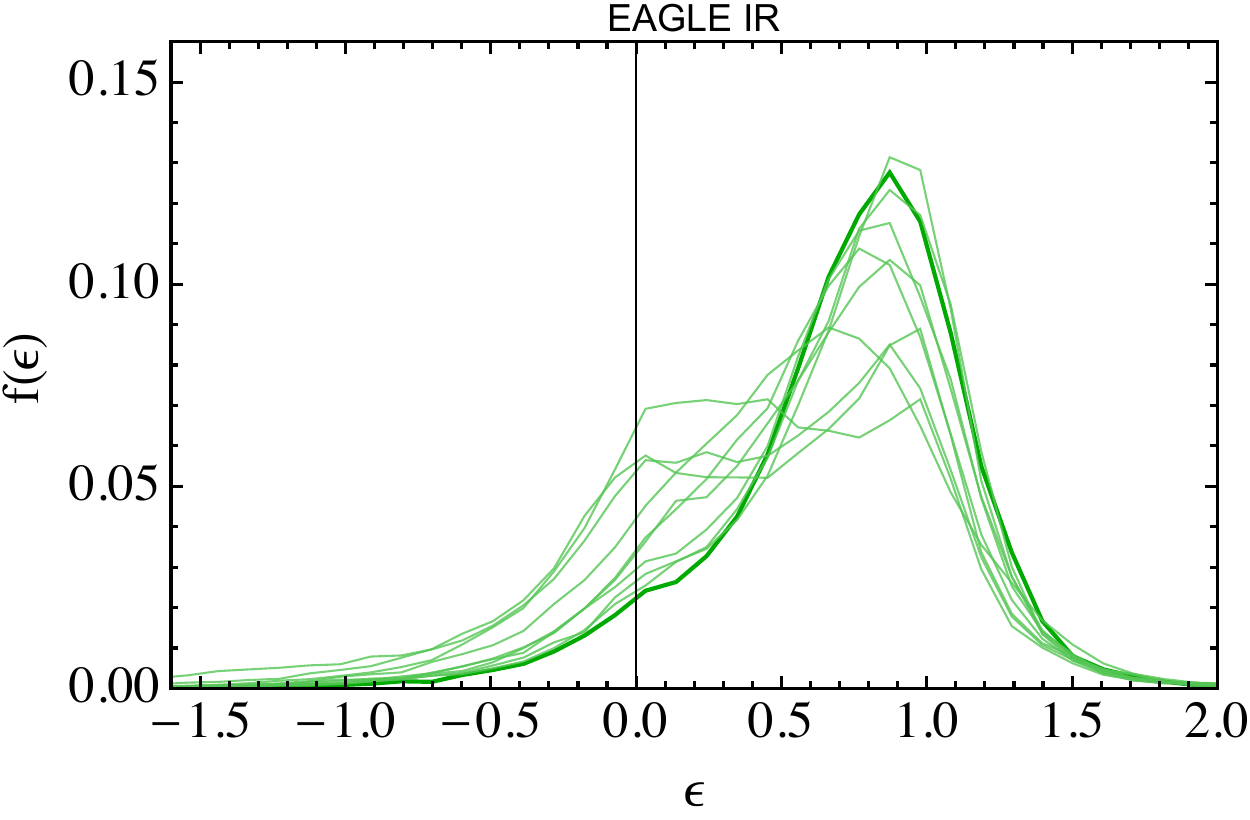}
        \includegraphics[width=0.48\linewidth]{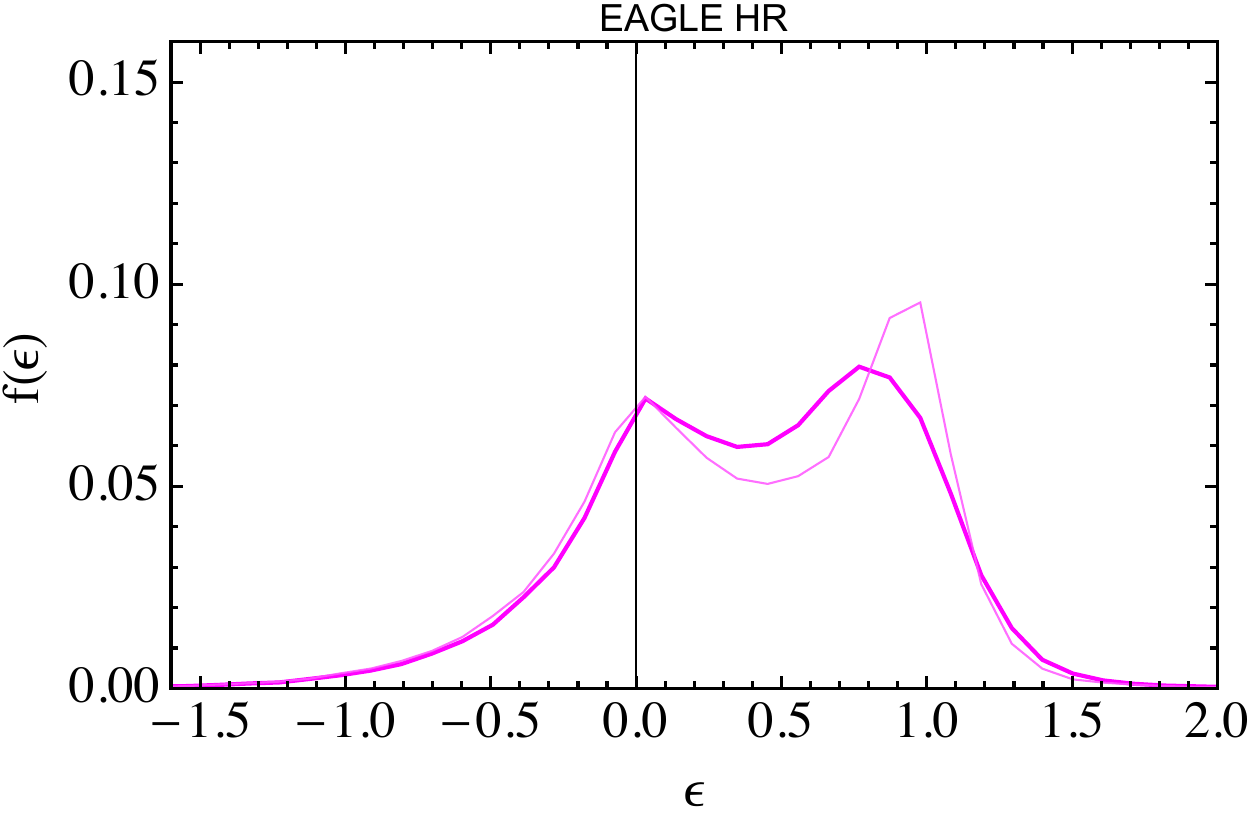} \\
         \includegraphics[width=0.48\linewidth]{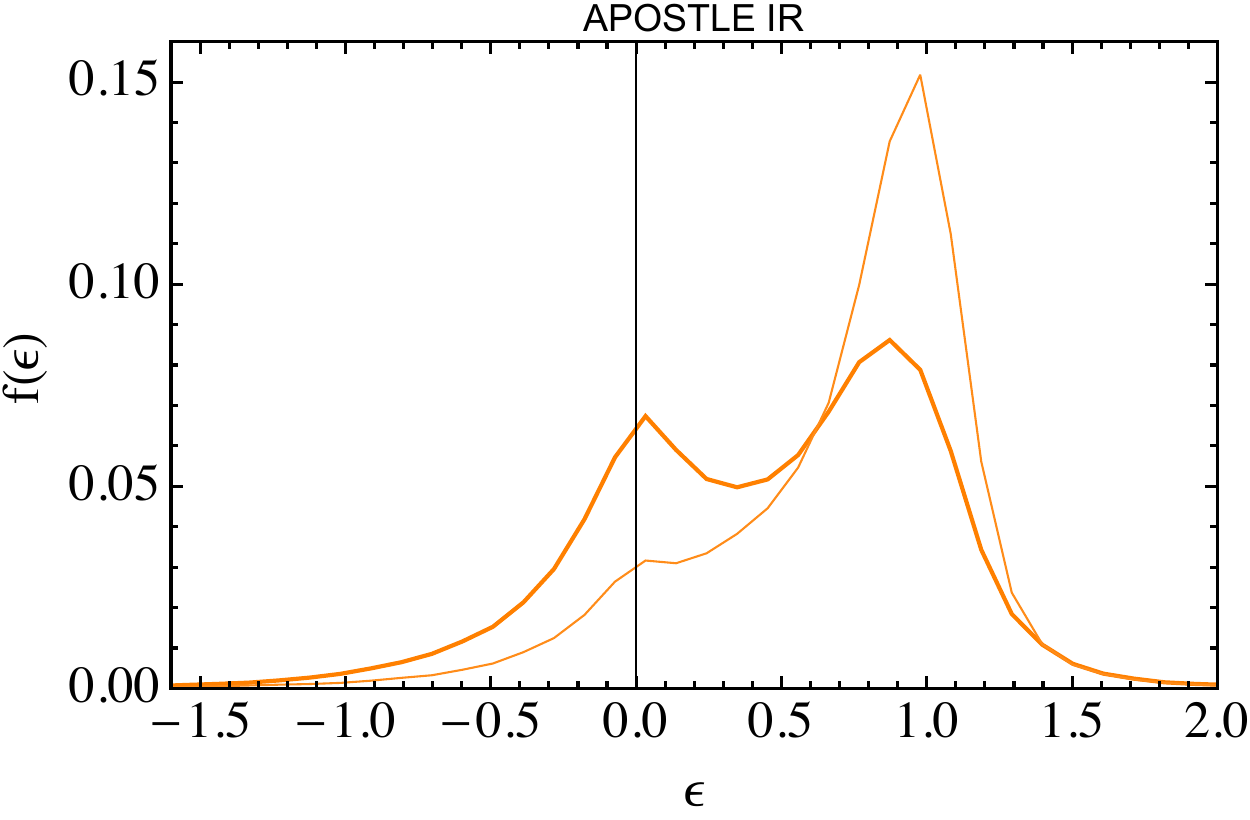} \\
    \end{center}
    \caption{Distribution of stellar circularities, $f(\epsilon)$, for the 
      haloes selected  in the EAGLE IR (\emph{top left} panel), EAGLE HR (\emph{top right} panel), and APOSTLE IR (\emph{bottom centre} panel)  resolution runs according to our criteria, including the requirement of having
      a significant disc stellar component. \emph{Thick} lines refer to the haloes
      giving the best-fit to the Galactic rotation curve for the three runs.}
    \label{fig:fepsilon}
\end{figure}

\begin{table}[t!]
    \centering
    \begin{tabular}{|c|c|c|c|c|c|c|}
      \hline
        Name & $M_{*}   [ \times 10^{10} \, \Msun]$ & $M_{200} [ \times 10^{12} \,  \Msun]$  &  D/T & $\chi^2/(N-1)$ \\
      \hline
   EAGLE IR   & 6.69 & 4.38 & 0.70 & 16.56\\
& 6.95 & 2.39  & 0.61 & 17.32\\
& 7.16 & 9.12  & 0.69 & 17.59\\
& 6.10 & 2.61  & 0.51 & 17.67\\
& 7.91 & 6.28 & 0.52 & 18.23\\
& 5.53 & 5.36 & 0.50 & 18.30\\
& 6.78 & 2.33  & 0.60 & 18.47\\
& 6.22 & 2.23  & 0.69 & 18.83\\
& 5.50 & 3.09  & 0.65  & 18.84\\
& 6.83 & 3.08  & 0.43 & 19.45\\
       \hline
      EAGLE  HR & 5.31 & 3.50 &  0.45 & 74.55 \\
& 5.48 & 2.76 & 0.46 & 220.96 \\
       \hline
       APOSTLE IR & 4.48 & 2.15 &  0.50 & 51.04 \\
& 4.88 & 1.68 &  0.70 & 221.27  \\
      \hline
    \end{tabular}
    \caption{Relevant parameters of the finally selected MW-like galaxies that satisfy
      our selection criteria in the EAGLE IR, EAGLE HR and APOSTLE IR runs.  
      We quote: total stellar mass,
      halo mass, disc-to-total mass ratio
      ($D/T$), and reduced $\chi^2$ for the fit to the rotation curve data. }
    \label{tab:selection}
  \end{table}
The MW is a spiral galaxy, with a well defined, unperturbed thin disc, a small bulge
and a bar component. We therefore seek to select simulated galaxies that are
themselves disc galaxies rather than ellipticals or undergoing mergers.

Following the approach of ref.~\cite{Scannapieco:2008cm}, we characterise the
dynamics of each simulated galaxy by looking for evidence of coherent rotation.
Each star particle in the simulated galaxies possesses an angular momentum vector
relative to its host's standard of rest. Any bulk stellar component that has the same
angular momentum vector as that of the hosting simulated galaxy will be considered to belong to the
disc. We can therefore use the distribution of angular momentum vectors of individual
particles relative to the net angular momentum of the galaxy to discriminate between
\emph{discs} (coherent rotation) and \emph{spheroids} (no coherent rotation;
comprises bulges and stellar haloes).

For each selected galaxy, we derive the distribution of the stellar orbital circularity parameter, $\epsilon$:
\begin{equation}
\epsilon(r) \equiv \frac{j_z}{j_c(r)} = \frac{j_z}{r v_c(r)},
\end{equation}
where $j_z$ is the component of the specific angular momentum parallel to the total
angular momentum of the galaxy, and $j_c(r)$ is the total specific angular momentum
of a circular orbit at distance $r$; here specific angular momentum is defined as
angular momentum divided by the star particle mass. The distribution of stellar
\emph{circularities}, $f(\epsilon)$, is a good indicator of the relative importance
of the disc with respect to the spheroidal stellar component. A disc in rotational
support, that is a configuration in which gravitational collapse is offset by the
centripetal acceleration, corresponds to a distribution peaked at about $\epsilon =
1$, while stars in a system supported by dispersion 
(i.e. a spheroidal system) show an almost symmetric distribution around $\epsilon =
0$~\cite{Scannapieco:2008cm,Scannapieco:2011yd,Sales:2012}.

As a further constraint -- in addition to the goodness of fit to the observed MW rotation
curve -- we want to select objects whose stellar kinematics shows a disc component
and, thus, are not completely dominated by the spheroidal contribution. When building
$f(\epsilon)$, we calculate the net specific angular momentum of the disc using only
those star particles with galactocentric radii in the range 3 -- 20 kpc. In this way,
our determination of the disc angular momentum direction should neither be affected by
the isotropic motions of bulge stars at low radii nor by high angular momentum
halo stars at large radii. We inspect the
distribution of stellar \emph{circularities}, $f(\epsilon)$, for the haloes giving
the best fit to the rotation curve data with the aim of building a final subset of
objects that pass our criteria and can be considered to be MW-like in our
perspective.  We retain a galaxy if the stellar fraction in the range $\epsilon > 0.45$
is larger than 50\%.  This criterion is meant to identify galaxies
that have a dominant disc and, on the other hand, to remove galaxies that show an
almost symmetric distribution around $\epsilon = 0$ -- and can thus be classified to
have a spheroidal only component. We tested different cuts on $f(\epsilon)$ and
adopted the one that is most conservative in retaining objects with a significant
disc component.  We note that the above criterion relies on the stellar kinematics
only, and that it might differ from the bulge-to-disc decomposition given by photometric
measurements. Indeed, photometric observations depend on assumptions on the shape of
the brightness profile of the spheroidal and disc
components~\cite{Scannapieco:2011yd,Scannapieco:2010nw}.  However, a full comparison of the
simulation outcome and the photometry of the MW is beyond the scope of the present
work and it would be interesting to study with follow-up analyses.

\medskip

In the case of the EAGLE IR run the number of galaxies passing the stellar mass \emph{and}
disc dominance criteria  is 145. Therefore, we further reduce the set of objects
of the EAGLE IR resolution run by selecting those that have a reduced $\chi^2 < 20$ for the fit 
to the MW rotation curve data.  
For the EAGLE IR run, we thus finally select  10 MW-like galaxies.
For the EAGLE HR resolution run, instead, we do not impose any further constraint on the
$\chi^2$, since only 2 objects survive the $f(\epsilon)$ criterion (and lie in the
correct stellar mass range), due to the 64 times smaller simulation volume. 
Finally, for the APOSTLE IR run, the 2 galaxies satisfying the 
dynamical and stellar mass constraints also match the disc requirement.

In figure~\ref{fig:fepsilon}, we show the distribution of stellar fraction for the
galaxies that give the best $\chi^2$ to the rotation curve data, have the acceptable MW
stellar mass, and show a substantial stellar fraction in the disc in the EAGLE IR, EAGLE HR, and APOSTLE IR
runs. 
The reduced $\chi^2$ values are quoted in table~\ref{tab:selection}. We note that in the cases of
the EAGLE HR and APOSTLE IR runs, the statistics for the initial sample are quite poor. In these cases the galaxy
with the lowest $\chi^2$ does not have any privileged statistical meaning, since the
true minimum of the distribution is not guaranteed to have been found because of the
small statistics. However, from the EAGLE IR results we are confident that the global minimum 
should lie within the correct MW stellar mass range. The fact that the $\chi^2$ values of
 the EAGLE HR and APOSTLE IR runs are worse
than the EAGLE IR one is thus not surprising. In what follows, we will use the best-fit galaxy as a reference
for illustrating the implications for DM indirect detection.

\smallskip

For the sake of completeness, we also quote here the reduced $\chi^2$ of the 4
galaxies belonging to the APOSTLE HR which are analysed in detail
in ref.~\cite{Schaller:2015mua}. The reduced $\chi^2$ values are (in order): 1203, 1982, 2056,
3554. Those are clearly poorer than the runs analysed in this work, but, again, the sample is
very limited in number. Moreover, we have checked that those galaxies have a lower
total stellar mass than what is observed within 3$\sigma$ ($M_{*}/(10^{10} \,
\Msun)=2.22$, 1.56, 1.05, and 1.06, from smallest to largest $\chi^2$
respectively). Since the stellar mass plays a role in the determination of rotation
curves in the inner 5 kpc, such a discrepancy in the stellar mass can be a further
reason for the poor $\chi^2$. 

\smallskip
 
\begin{figure}
    \begin{center}
        \includegraphics[width=0.21\linewidth]{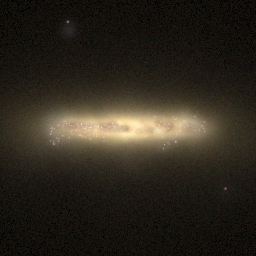}
        \includegraphics[width=0.21\linewidth]{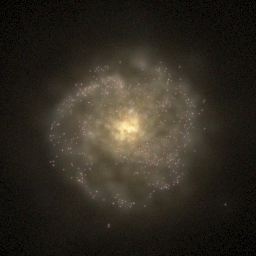}
         \hspace{10pt} \includegraphics[width=0.21\linewidth]{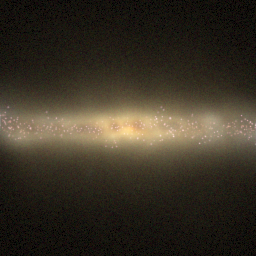}
        \includegraphics[width=0.21\linewidth]{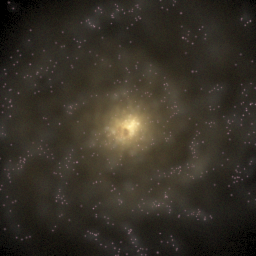}\\ \vspace{3pt}
        \includegraphics[width=0.21\linewidth]{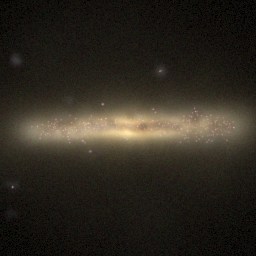}
        \includegraphics[width=0.21\linewidth]{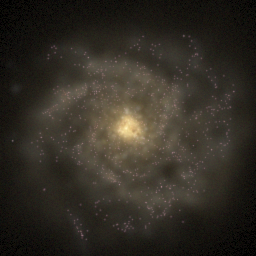}
       \hspace{10pt} \includegraphics[width=0.21\linewidth]{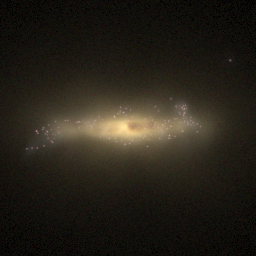}
        \includegraphics[width=0.21\linewidth]{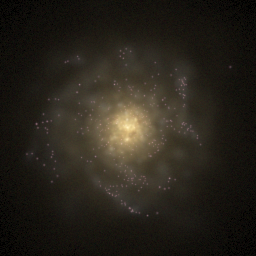}\\ \vspace{3pt}
        \includegraphics[width=0.21\linewidth]{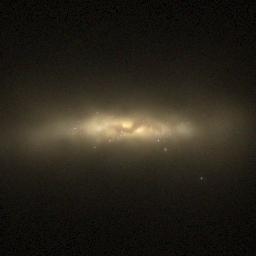}
        \includegraphics[width=0.21\linewidth]{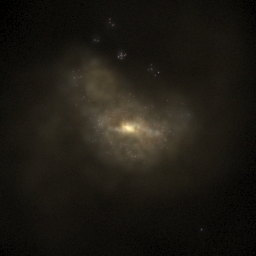}
       \hspace{10pt} \includegraphics[width=0.21\linewidth]{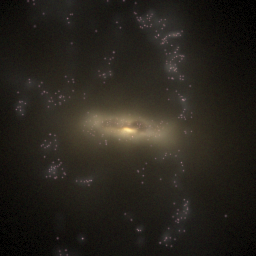}
        \includegraphics[width=0.21\linewidth]{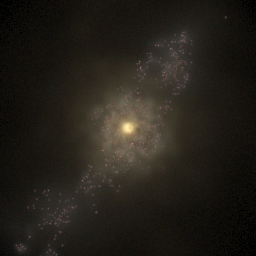}\\ \vspace{3pt}
        \includegraphics[width=0.21\linewidth]{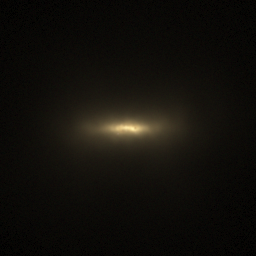}
        \includegraphics[width=0.21\linewidth]{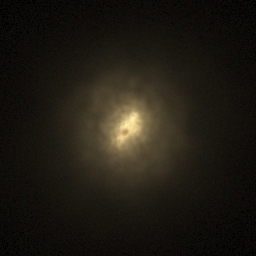}
       \hspace{10pt} \includegraphics[width=0.21\linewidth]{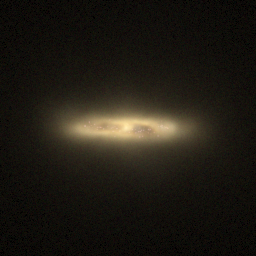}
        \includegraphics[width=0.21\linewidth]{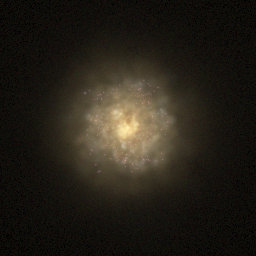}\\ \vspace{3pt}
        \includegraphics[width=0.21\linewidth]{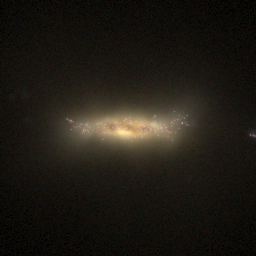}
        \includegraphics[width=0.21\linewidth]{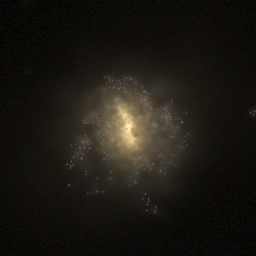}
       \hspace{10pt} \includegraphics[width=0.21\linewidth]{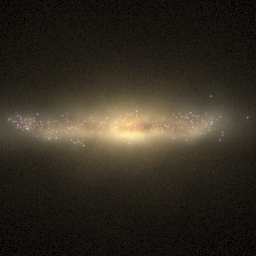}
        \includegraphics[width=0.21\linewidth]{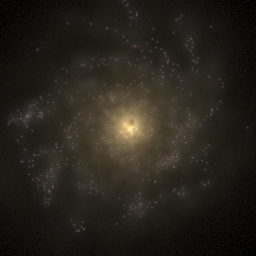}
    \end{center}
    \caption{Pairs of mock three-colour composite grid observations for the 10
      simulated galaxies in the EAGLE IR run with edge-on (\emph{left}) and face-on
      (\emph{right}) projections, within a 30 kpc aperture of the galaxy centre
      obtained using the \textsc{skirt} \cite{Camps:2015} radiative transfer code as
      described by \cite{2015MNRAS.452.2879T,Trayford15}.  }
    \label{fig:images}
\end{figure}

\begin{figure}
    \begin{center}
        \includegraphics[width=0.29\linewidth]{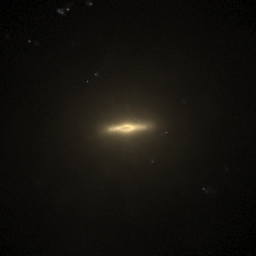} \hspace{5pt} 
        \includegraphics[width=0.29\linewidth]{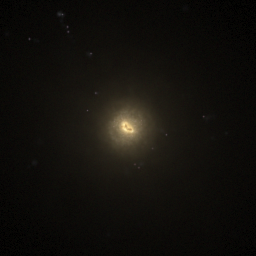}\\
        \vspace{5pt} 
        \includegraphics[width=0.29\linewidth]{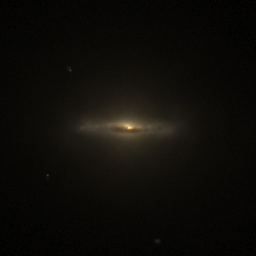} \hspace{5pt} 
        \includegraphics[width=0.29\linewidth]{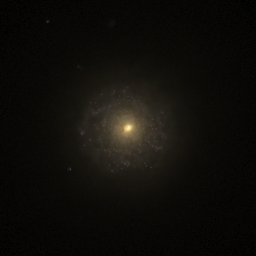}
    \end{center}
    \caption{Same as figure~\ref{fig:images} for the 2 selected MW-like galaxies in the EAGLE
      HR run.}
    \label{fig:images2}
\end{figure}

\begin{figure}
    \begin{center}
        \includegraphics[width=0.30\linewidth]{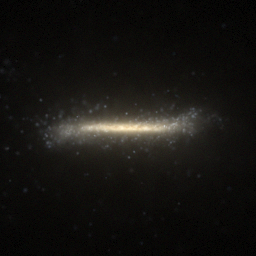} \hspace{5pt} 
        \includegraphics[width=0.30\linewidth]{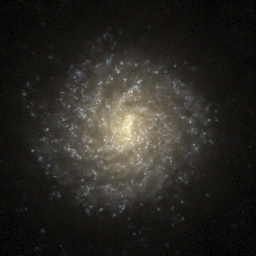}\\
        \vspace{5pt} 
        \includegraphics[width=0.30\linewidth]{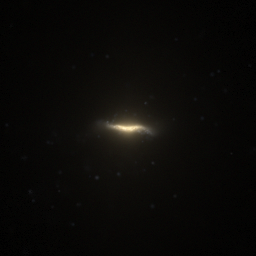} \hspace{5pt} 
        \includegraphics[width=0.30\linewidth]{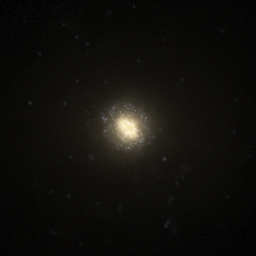}
    \end{center}
    \caption{Same as figure~\ref{fig:images} for the 2 selected MW-like galaxies in the
      APOSTLE IR run.}
    \label{fig:images3}
\end{figure}

To corroborate the implementation of criterion (iii), in
figures~\ref{fig:images}, \ref{fig:images2} and~\ref{fig:images3} 
we display the images of the edge-on and face-on orientations of the selected MW-like galaxies in
the EAGLE IR, EAGLE HR and APOSTLE IR run.  
The images are produced with the radiative transfer code {\sc skirt} \cite{Camps:2015}.  
This code generates images of the galaxy in the {\emph u, g}, and {\emph r} SDSS 
filters within a 30 kpc aperture of the galaxy centre; the dust distribution is inferred from the gas metal
distribution in the interstellar medium~\cite{2015MNRAS.452.2879T,Trayford15}.  From
those images one can see that the majority of the selected objects shows clearly a
disc component.

\smallskip 

To further test the prominence of the disc, we compute the disc-to-total mass ratio
for the set of selected MW-like galaxies. The disc-to-total mass
ratio writes as~\cite{Scannapieco:2008cm},
\begin{equation}
\frac{D}{T} \equiv \frac{M_d}{M_d + M_s} \, ,
\end{equation}
where $M_d = \sum_{\epsilon_i > 0.6} (f(\epsilon_i) \, m_g)$ is the stellar mass in
the disc, and $M_s = \sum_{\epsilon_i > -0.6}^{\epsilon_i < 0.6} (f(\epsilon_i) \,
m_g)$ is the stellar mass in the spheroid.  The disc-to-total mass ratios are listed
in table~\ref{tab:selection} for the sets of the selected MW analogues in the three
simulations and are all in the range 0.4 -- 0.7.  For the EAGLE HR simulation run, the
ratios are systematically lower than what is expected for real spiral
galaxies~\cite{Graham:2008hn}, while for some galaxies in the EAGLE IR and APOSTLE IR runs the D/T ratio is closer to
observations of the MW, $(D/T)_{\rm MW} \sim 0.86$~\cite{McMillan:2011wd}. 

\bigskip

In summary, we have identified a minimal set of 10, 2 and 2 galaxies for the EAGLE IR, EAGLE HR
and APOSTLE IR
simulation runs that simultaneously (i) give a good fit to the rotation curve of
the MW, (ii) have the right total stellar mass (within 3$\sigma$) of the measured
value, and (iii) show a significant disc stellar component. In
table~\ref{tab:selection}, we summarise the main properties of the chosen galaxies for
the three simulation runs.  These three subsets of objects represent MW analogues
according to our definition. We have thus demonstrated that minimal selection
criteria, such as the requirement to give a good fit to the observed rotation curve,
significantly reduce the number of prototypical MW-like galaxies.  In
sections~\ref{sec:rhodm} and~\ref{sec:idm} we will use the two sets of galaxies to
investigate the distribution of DM in the halo and discuss implications for indirect
DM searches.
We finally stress that discussing the matching 
of EAGLE simulated galaxies with other MW observables, such as the brightness emission
profile, is beyond the scope of the present paper. Further studies along this direction would be extremely helpful for the field.
 
\section{The Galactic dark matter density profile}  
\label{sec:rhodm}

\begin{figure}[t!]
    \begin{center}
        \includegraphics[width=0.49\linewidth]{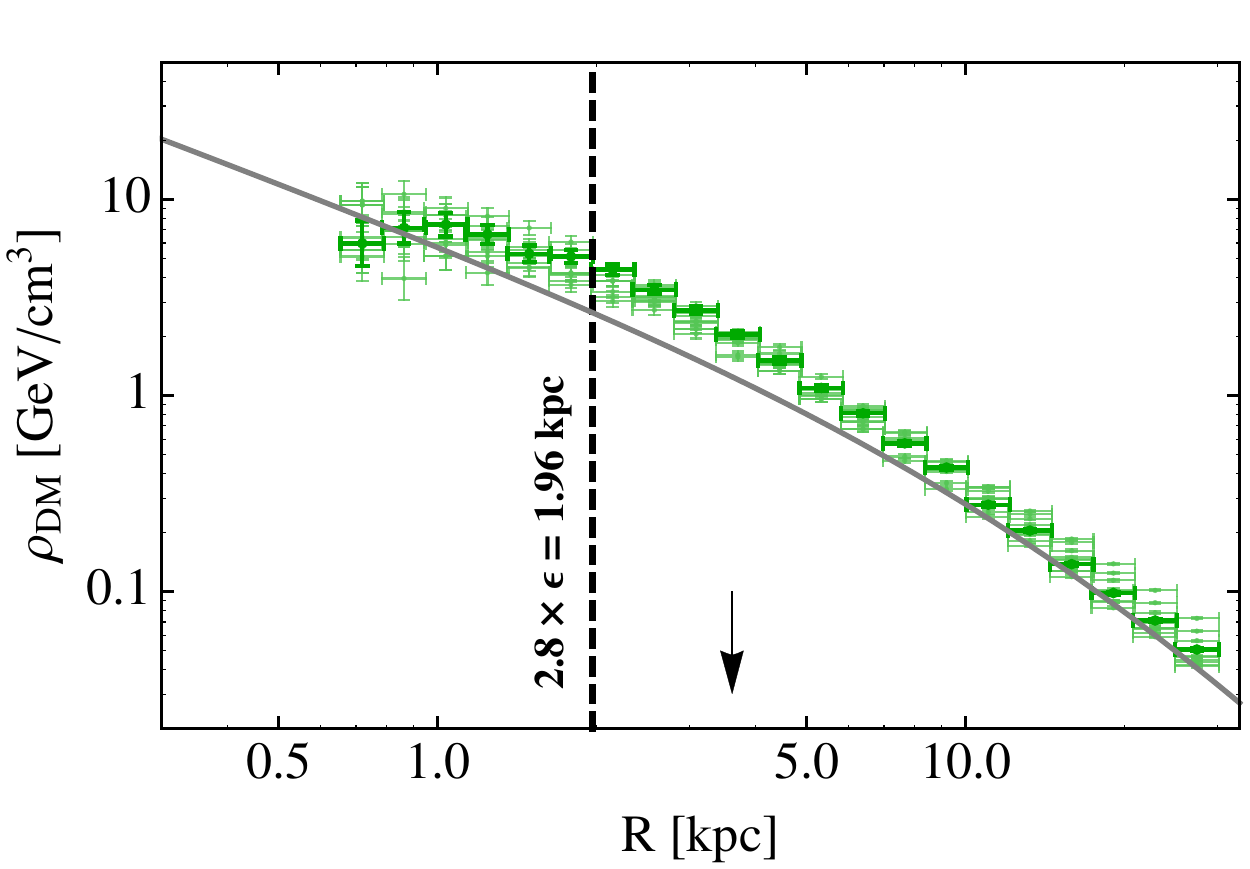}
        \includegraphics[width=0.49\linewidth]{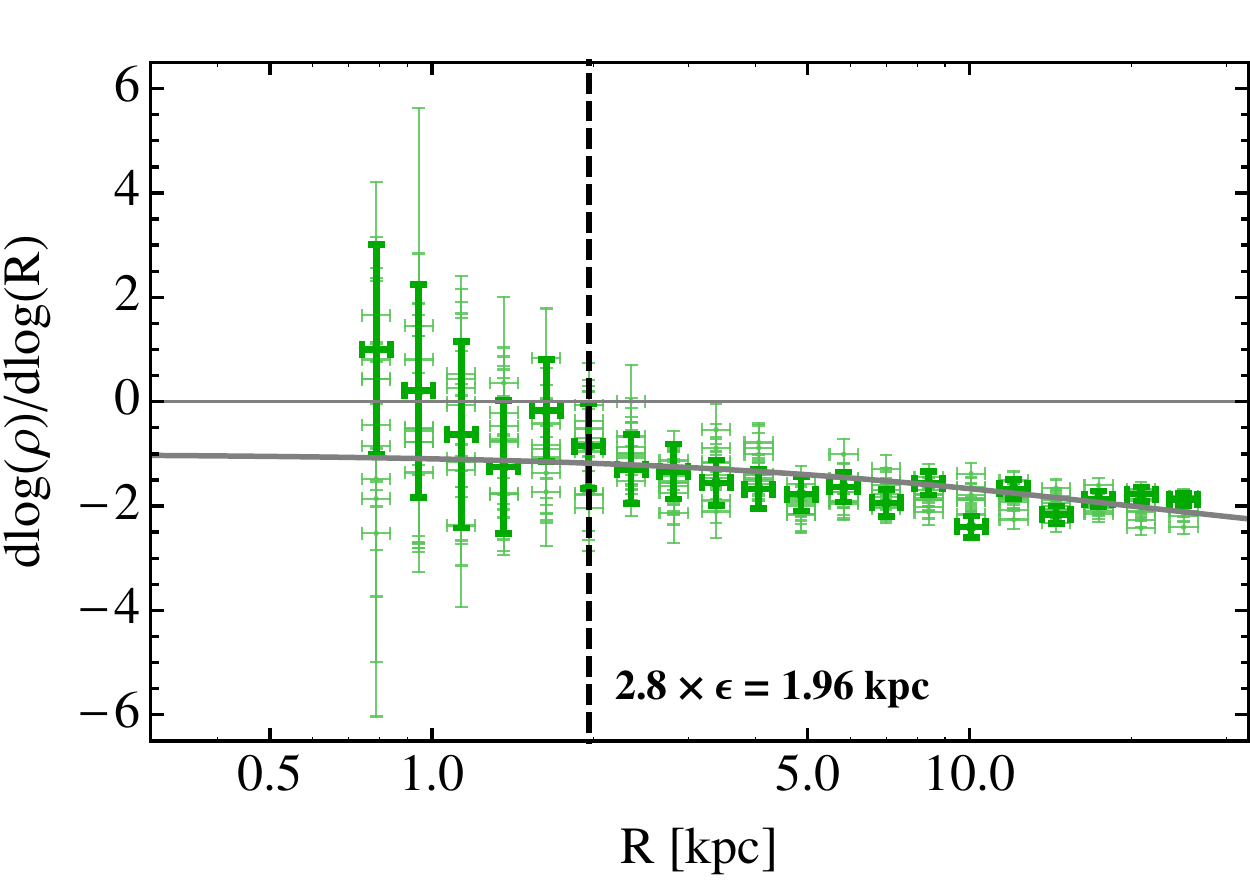}\\
        \includegraphics[width=0.49\linewidth]{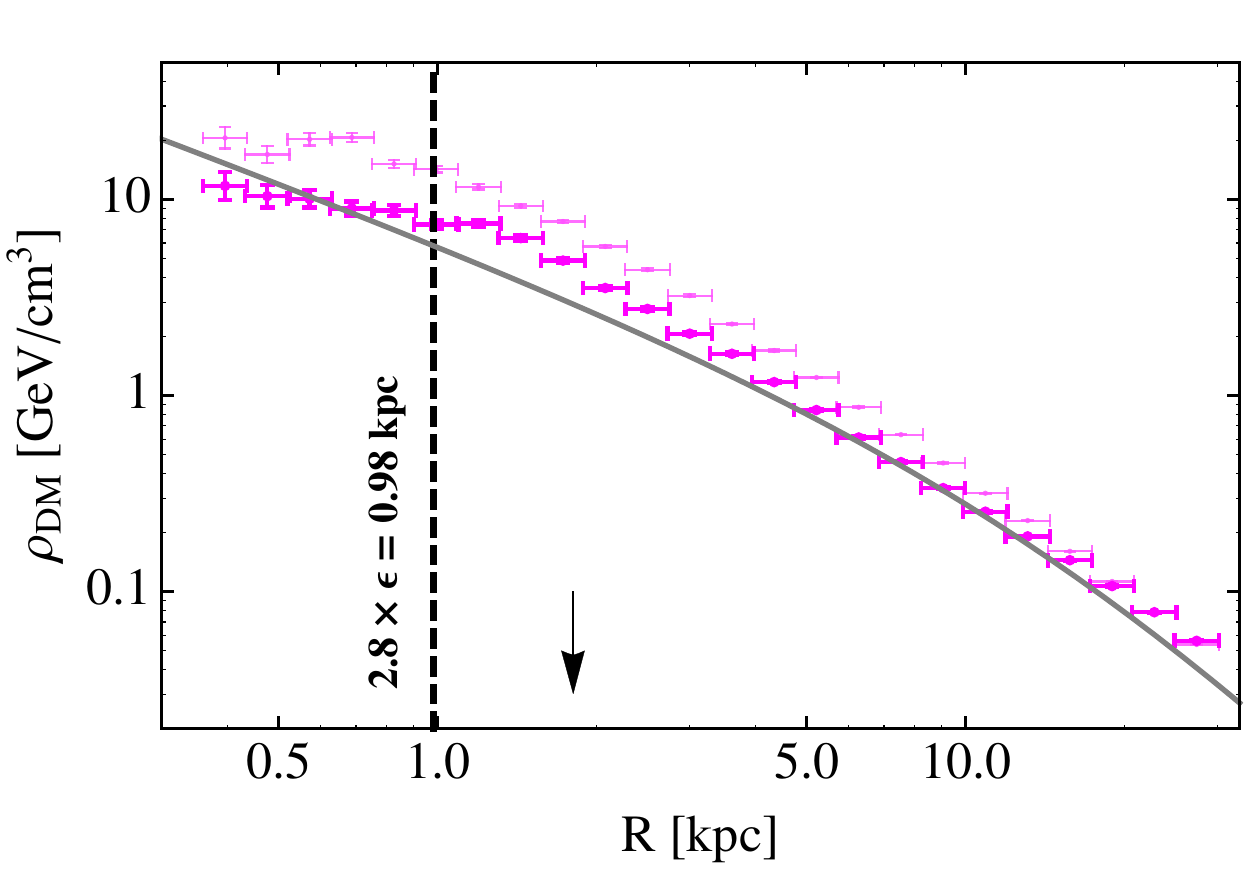}
        \includegraphics[width=0.49\linewidth]{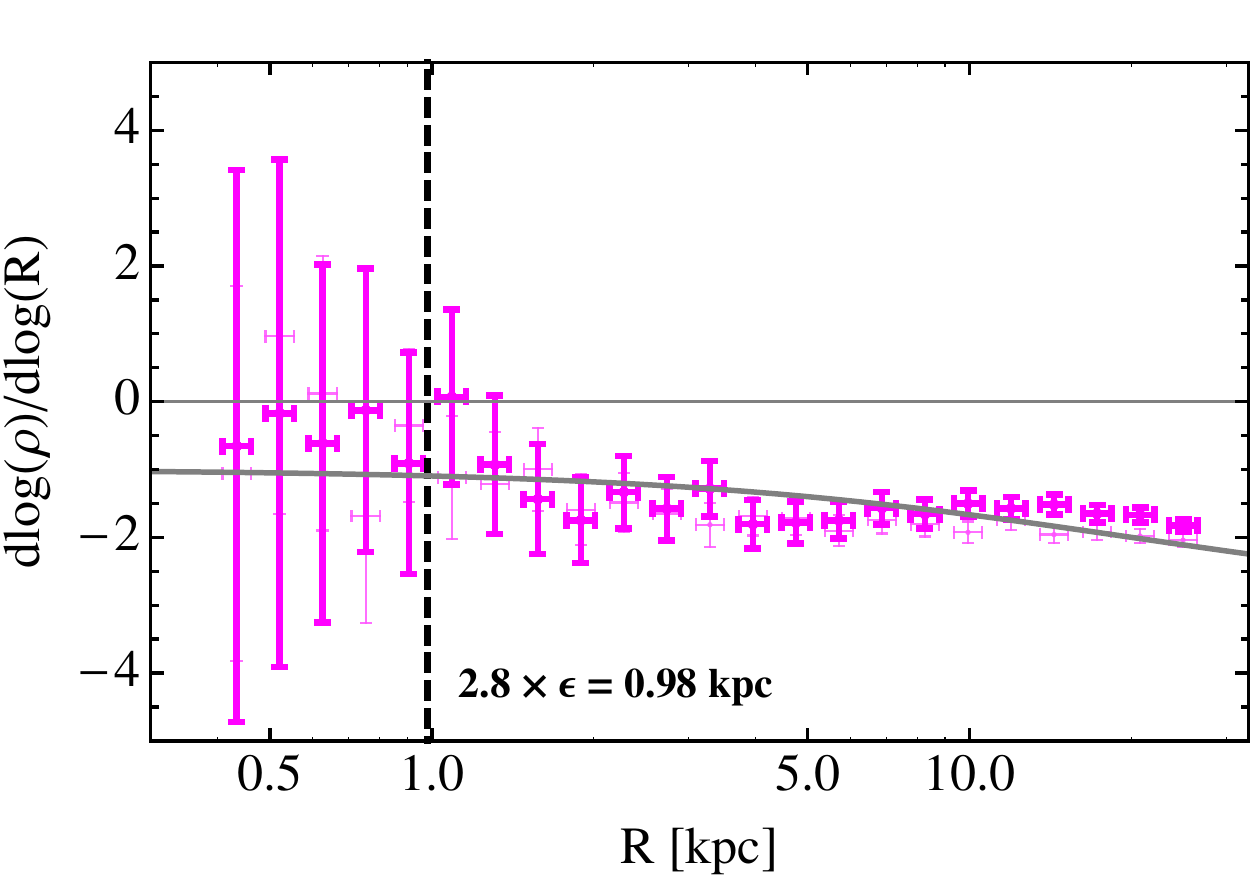}\\
        \includegraphics[width=0.49\linewidth]{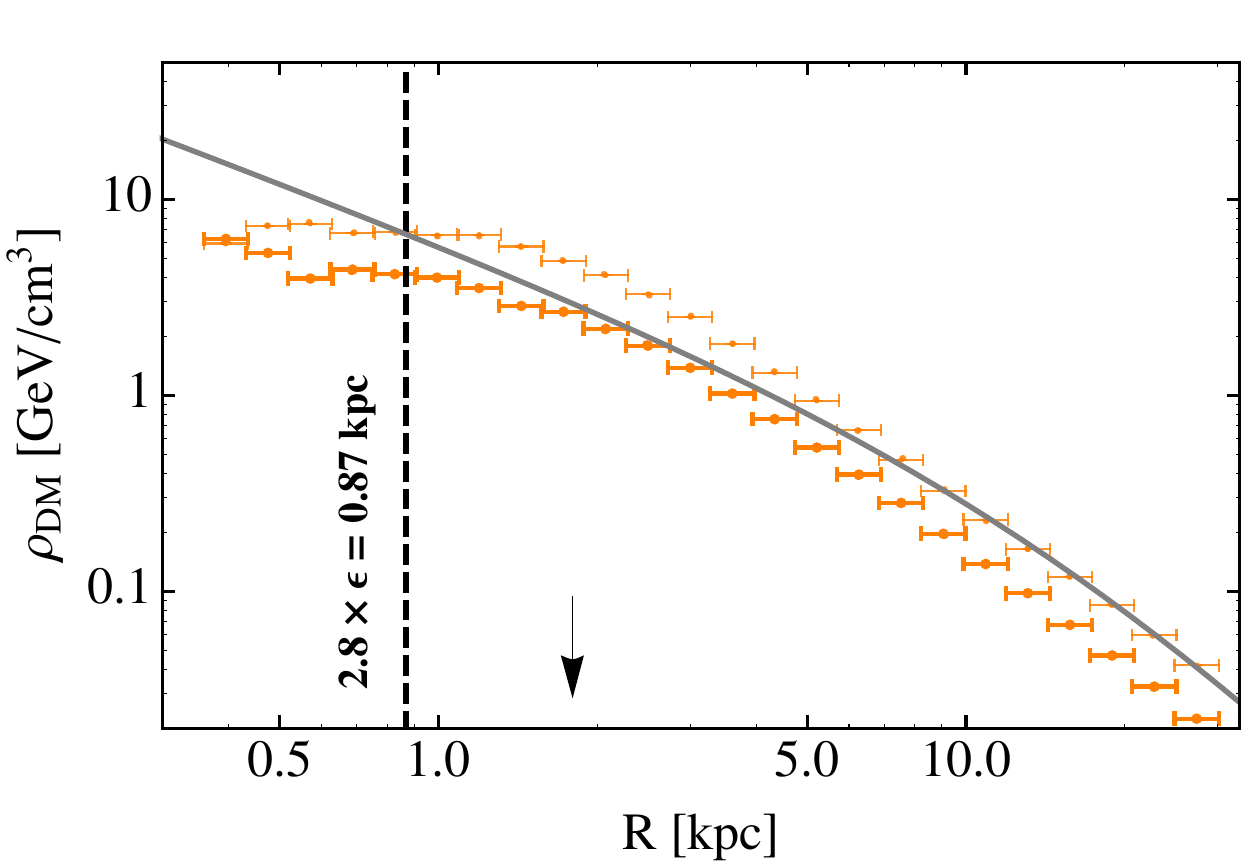}
       \includegraphics[width=0.49\linewidth]{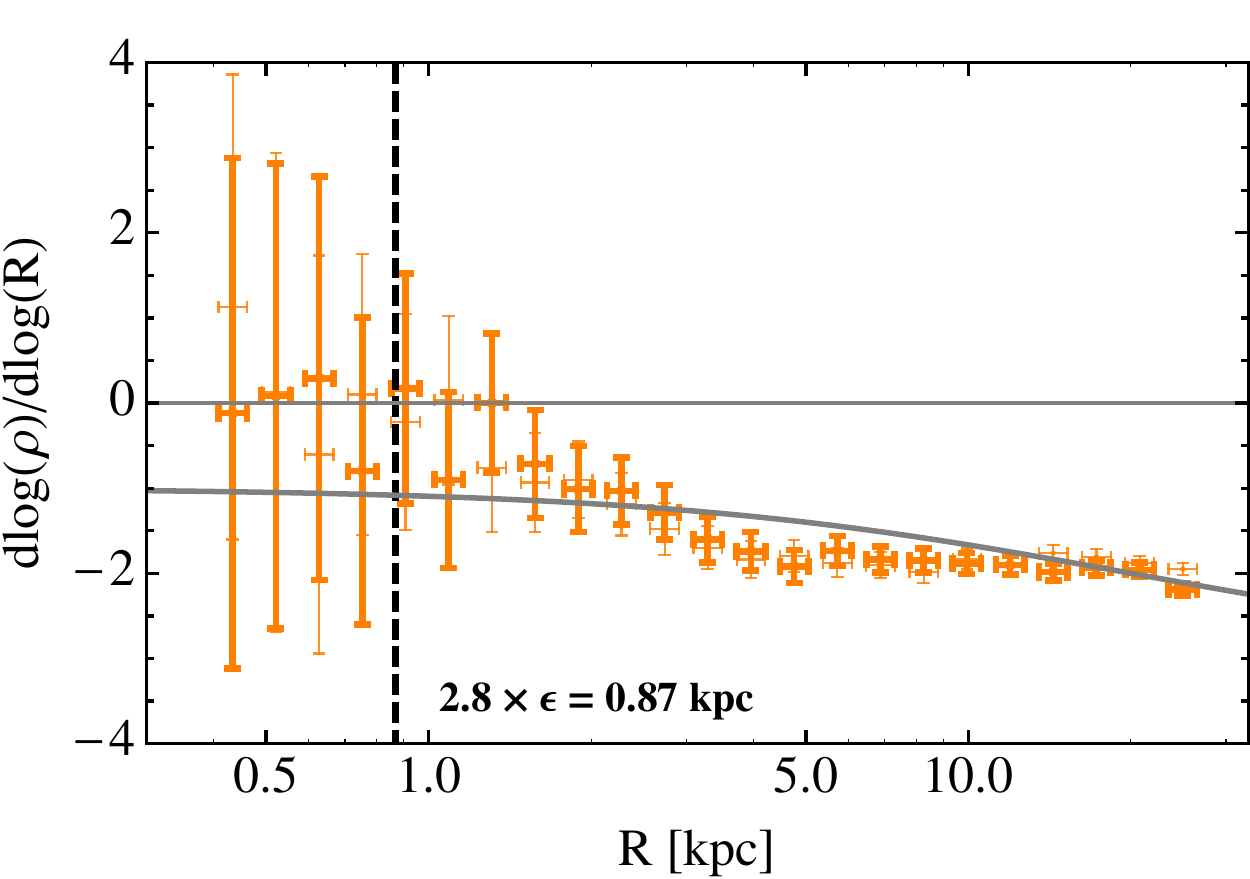}
    \end{center}
    \caption{DM density profiles (\emph{left} panels) and the radial change of the
      local logarithmic slopes (\emph{right} panels) of the selected MW-like galaxies in
      the EAGLE IR (\emph{top}), EAGLE HR (\emph{middle}) and APOSTLE IR
      (\emph{bottom}) runs. The \emph{thick grey line}
      represents the prediction for an NFW profile with $r_s = 20$ kpc and local DM
      density $\rho_\odot = 0.4$ GeV/cm$^3$ (as commonly assumed in DM indirect detection
      studies).  In all panels the effective resolution
      of the simulation is shown by the \emph{dashed black line}, while the \emph{black
      arrows} on the left panels indicate the convergence radii of 3.6 kpc (EAGLE IR) 
      and 1.8 kpc (EAGLE HR and APOSTLE IR) as discussed in the text. }
    \label{fig:profile_IR_HR}
\end{figure}

In this section we analyse the DM density profile of the final set of MW-like galaxies
and discuss possible implications for DM indirect detection.  As explained in
section~\ref{sec:intro}, the predicted DM annihilation flux is very sensitive to the
DM density profile and it is thus important to study the predictions of hydrodynamic
simulations and any deviations from pure DM N-body simulation density profiles.  As for
the EAGLE simulations, it has been shown that, on average, haloes in the MW mass
range ($\sim 10^{12} \Msun$) have a DM density profile that significantly deviates
from the NFW profile and has a core in the inner few
kiloparsecs~\cite{Schaller:2014uwa}.  Although the flattening might appear below the
effective resolution (depending on the resolution run considered), the presence of a
core on small radial distances from the GC seems to be a universal feature of EAGLE
simulated objects.  This has been recently demonstrated for the set of galaxies of
the APOSTLE project, which were run with the same code and subgrid physics
model as EAGLE; the density profile properties of the highest APOSTLE simulation run have been analysed in detail
in~\cite{Schaller:2015mua}.  We turn now to the analysis of the density profiles of our
final set of MW-like galaxies.

\medskip

In the left panels of figure~\ref{fig:profile_IR_HR}, we show the DM density profile
of the final set of haloes for the EAGLE IR, EAGLE HR  and APOSTLE IR runs. The 
DM density is derived directly
from the mass enclosed in a given shell between R and R + $\delta R$, adopting
equally spaced radial bins in logarithmic space.  By construction, we assume
spherical symmetry for the distribution of DM, which has been shown to be a
good assumption for the APOSTLE simulations~\cite{Schaller:2015mua}. Furthermore,
it has been shown that baryons tend to make the distribution of DM more
spherical than in simulations including solely DM
\cite{Dubinski:1994,Bryan:2013,Bernal:2014mmt}. We verified that for our set of MW-like galaxies the
assumption of a spherically symmetric profile is a good approximation.
The uncertainty in the density is
given by the Poissonian error in the number of particles in each mass shell (the
error in the distance refers instead to the adopted binning).  As presented in
section~\ref{sec:simulation}, the effective resolution limit of the simulation runs,
i.e. the Plummer-equivalent softening length, is $2.8 \times \epsilon$, where $\epsilon=0.7~\rm{kpc}, 0.35~\rm{kpc}$
and $0.31~\rm{kpc}$ for the EAGLE IR, EAGLE HR and APOSTLE IR runs, respectively. However, the radius
at which profiles can be considered as converged is larger than this value and can be
estimated in collisionless simulations using the criterion of
\cite{Power:2002sw} that identifies the radius at which the integral in mass is independent on the resolution. 
The so-called ``Power radius" is $R_{\rm P03} = 3.6~\rm{kpc}, 1.8~\rm{kpc}$
and $1.8~\rm{kpc}$\footnote{The Power radius is a halo-dependent quantity and thus it is defined
 for each halo. We checked however that the variation of $R_{\rm P03}$ among haloes of the same simulation
 run varies only within a few percent. Therefore, using an average value for $R_{\rm P03}$ is a good approximation.}
 for the EAGLE IR, EAGLE HR and APOSTLE IR runs.
 The very concept of convergence -- that ever increasing resolution will cause the measured 
quantity to asymptotically tend towards some finite value -- is much less clear in simulations
 that feature baryon physics, and so the innermost radius at which the profiles may be considered 
 converged is ill-defined. A discussion of these issues
can be found in refs.~\cite{Schaller:2014uwa,Schaye:2015,Schaller:2015mua}. 

In figure~\ref{fig:profile_IR_HR} we
show both the resolution limit and the Power radius for the three
resolution runs. Between those two radial scales the results of the simulation have to
be treated with extreme caution, as there might still be some residual numerical
effects due to discreteness and softening of the gravitational force. This is particularly true for the EAGLE IR run, 
while it is less dramatic for EAGLE HR and APOSTLE IR.

In general, the DM density is shallower than what is expected from an
NFW profile in the inner 1.5 -- 2 kpc.  Between about 1.5 -- 6/8 kpc, however, the
effect of baryons results in a steeping of the DM profile.  Analogous features have
been found in the APOSTLE HR simulations~\cite{Schaller:2015mua}.  The origin of
the flattening in~\cite{Schaller:2015mua} might be related to the star
formation history of the haloes: while they developed a cusp at redshifts larger than
1, at $z < 1$ episodes of sudden enhancement of the star formation rate
occurred. Such violent starbursts could have destroyed the cusp~\cite{Navarro:1996bv}
at sufficiently late times that the galaxy cannot subsequently accrete enough small DM
haloes to rebuild the cusp. 
We stress here that the mild flattening occurring in EAGLE IR between the resolution limit
and the Power radius is most likely due to softening effects and 
limitations in the gas physics model~\cite{Schaller:2015mua}, while for EAGLE HR
and APOSTLE IR it might be a combination of the softening and of physical effects.

To better appreciate the deviation of the simulated haloes' profiles from pure DM
simulations, i.e.~from an NFW density profile, in the right panels of
figure~\ref{fig:profile_IR_HR} we show the radial variation of the local logarithmic
slope for our selected MW-like galaxies together with the expectation for the NFW
density profile.  In the case of the EAGLE IR run, it is not possible to establish the
presence of the flattening since the error bars on the logarithmic slope make it
compatible with the NFW profile down to the resolution limit.  This is compatible
with what is found by~\cite{Schaller:2014uwa}, who studied properties of stacked
haloes in the EAGLE IR and EAGLE HR simulations. However, for the EAGLE HR and
APOSTLE IR selected MW-like galaxies,
there is a deviation of the slope from -1 and a tendency to 0 slightly
above the resolution limit.  From the same panels (EAGLE HR and APOSTLE IR runs) we can also appreciate the
effect of the baryonic contraction: in the range 1.5 -- 6 kpc the logarithmic slope
is steeper than -1.

\medskip

The two features of the DM profile (flattening below 1.5 kpc and adiabatic
contractions below 10 kpc) appear thus to be universal properties of the simulated MW-like galaxies within
the EAGLE galaxy formation model, i.e.~a generic outcome of the subgrid model assumed.
 What is striking, though not surprising, is that for the
selected MW-like galaxies the overall normalisation of the different DM haloes does not show
 large scatter. This is again a direct consequence of
demanding that the selected galaxies fit the kinematical data of the MW.  As a consequence, 
the variation in the local DM density is also small. 
For the final sets MW-like galaxies,
$\rho_\odot$ (with R$_\odot$ = 8.0 kpc) spans the range 0.44 -- 0.59 GeV/cm$^3$ (EAGLE IR),
0.41 -- 0.56 GeV/cm$^3$ (EAGLE HR), and 0.41 GeV/cm$^3$ (APOSTLE IR).
 This is compatible with recent measurement of the
local DM density, whose standard value at 8.5 kpc is assumed to be 0.4 GeV/cm$^3$
(which translates to 0.44 GeV/cm$^3$ at 8.0 kpc)~\cite{Catena:2009mf,2010A&A...523A..83S}.  We acknowledge
however that the local DM density is affected by several uncertainties and, in
particular, the DM density has been found to be larger in the stellar disc compared
to what is derived from a spherical shell~\cite{Pato:2010yq}.

\section{The implications for dark matter indirect detection} 
\label{sec:idm}
\begin{table}
    \centering
    \begin{tabular}{|c|c|c|c|c|}
      \hline
       Profile & $\langle \sigma v \rangle  [ \times 10^{-26} \, \rm cm^{3}/s]$ & $m_{\chi} \, [\rm GeV]$ & $\chi^2$ & $p$-value \\
       \hline
       gNFW ($\gamma$=1.26) & $1.71 \pm 0.11$& $47.32 \pm 1.07$ & 223.9 & 0.73\\
      \hline
      EAGLE HR & $  1.96 \pm 0.14$ & $46.37 \pm 1.37 $ & 246.3 & 0.34   \\
       \hline
      APOSTLE IR & $ 1.76 \pm 0.16$ &  $45.36 \pm 2.96 $& 283.9 & 0.02  \\ 
      \hline
    \end{tabular}
    \caption{Best-fit parameters ($\langle \sigma v \rangle$ and mass) together with goodness of fit indicators 
    	($\chi^2$  -- with 238 degrees of freedom -- and $p$-value) of the ten sub-regions fit for the gNFW profile,
	 and the best-fit galaxies of the EAGLE HR and APOSTLE IR runs.
    	100\% annihilation into b-quarks is assumed. }
    \label{tab:svm}
  \end{table}

\begin{figure}
    \begin{center}
        \includegraphics[width=0.6\linewidth]{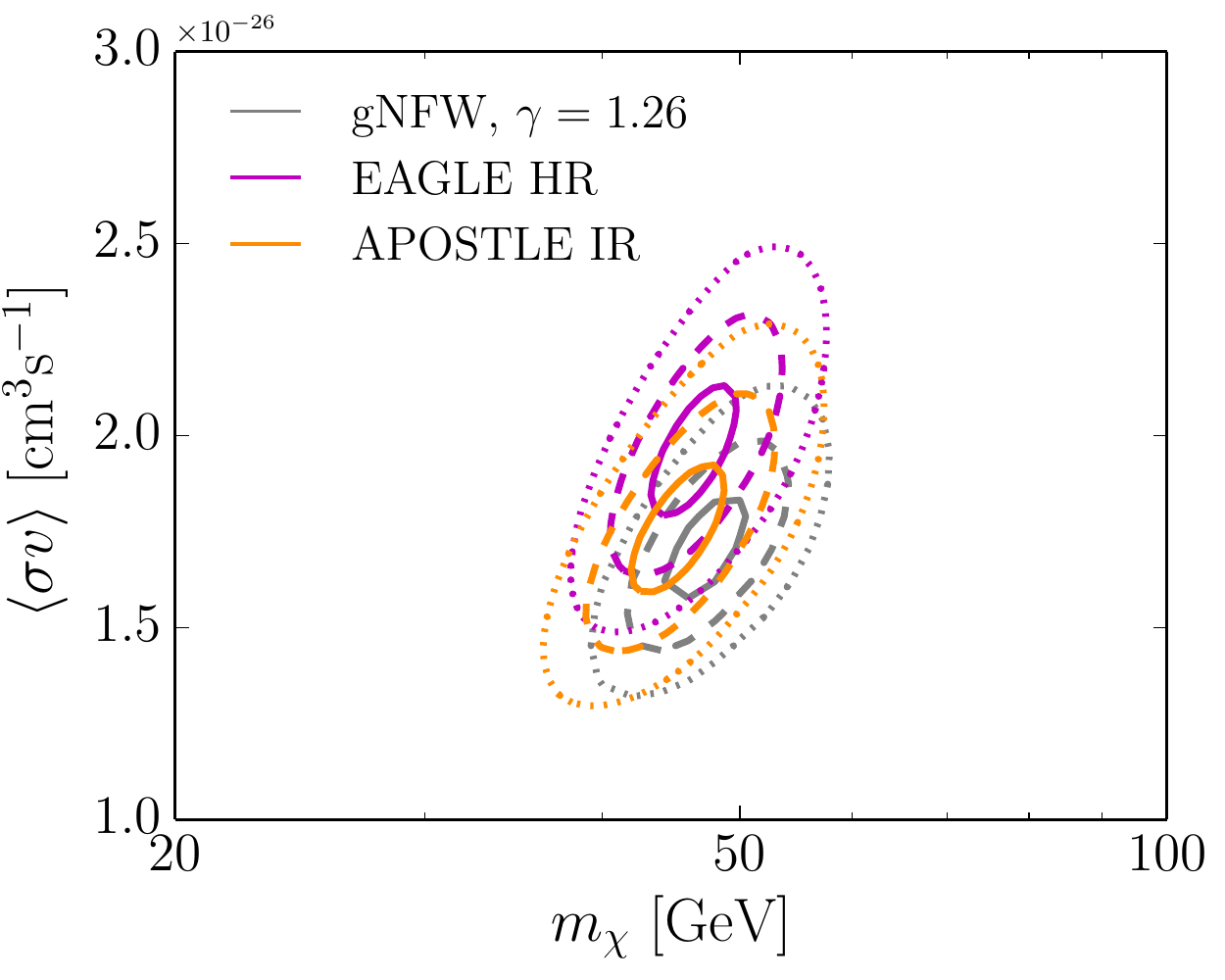}
    \end{center}
    \caption{Constraints on the ($\langle \sigma
    v\rangle$, $m_\chi$) plane for annihilation into b-quarks. The three contours refer
    to different DM density profiles: the EAGLE HR run best-fit MW-like galaxy's DM density profile (\emph{magenta}), 
    the APOSTLE IR one (\emph{orange}), and the gNFW profile with $\gamma=1.26$ (\emph{grey})
    (\cf table~\ref{tab:svm}).}
    \label{fig:contours}
\end{figure}

\begin{figure}[h!]
    \begin{center}
        \includegraphics[width=0.95\linewidth]{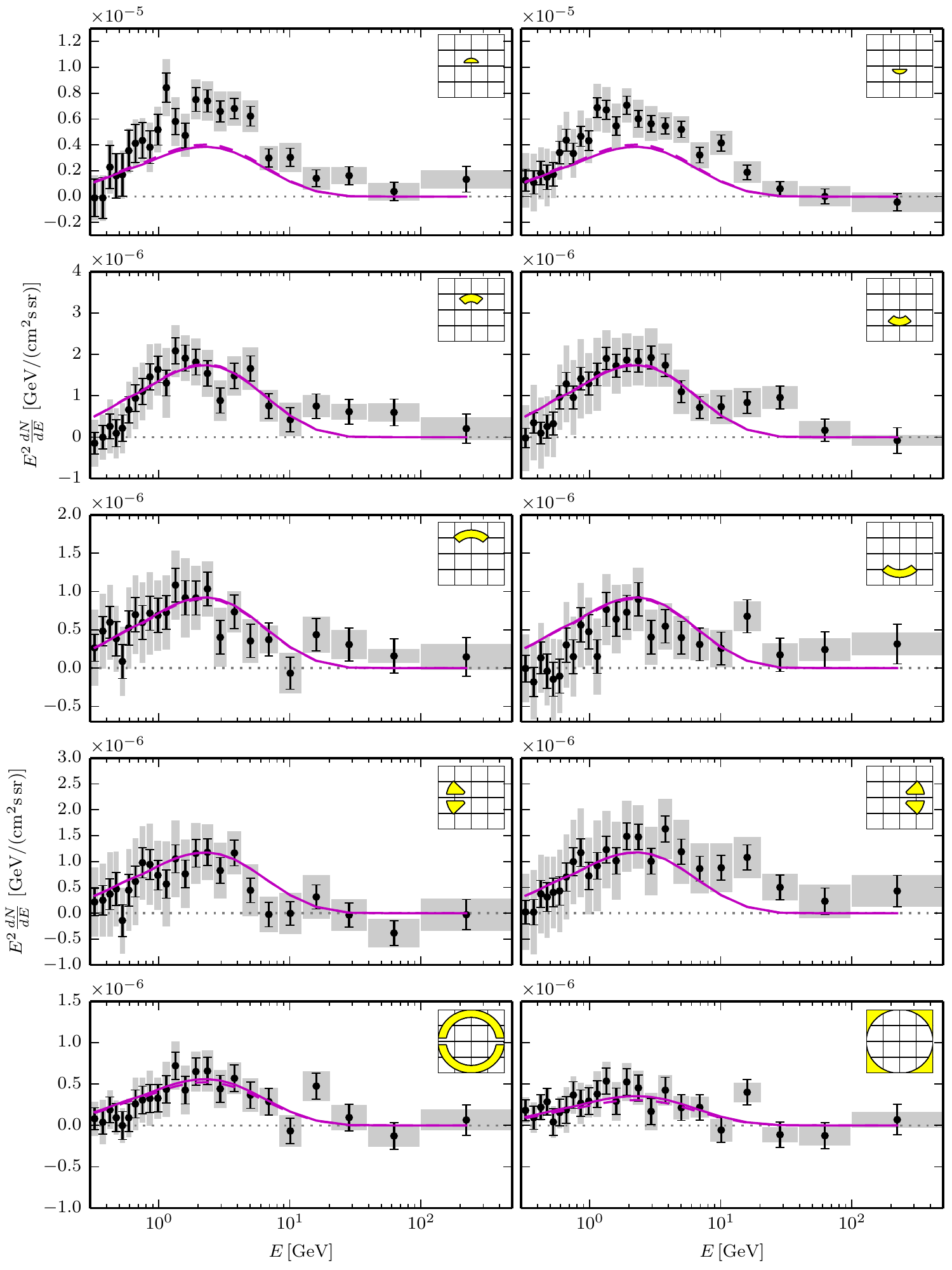}
    \end{center}
    \caption{Fluxes in the ten sub-regions for the two MW-like galaxies selected for the EAGLE HR run.
     	The \emph{solid} line indicates the best-fit galaxy, while 
	the \emph{dashed} line corresponds to the other galaxy in the final sample (in most of the regions
	the two lines overlap because of the fit results). For each galaxy, the flux is computed
	for mass and cross section as obtained by the fit to the ten sub-regions (see text for details). 
	The \emph{black points} are the GeV excess data as derived in~\cite{Calore:2014xka}, with corresponding systematic
	uncertainties (\emph{grey boxes}). The insets refer to the 
	region of interest over which the flux is averaged, see~\cite{Calore:2014xka} for the definition of the ten regions.
     }
    \label{fig:fluxes_HR}
\end{figure}

\begin{figure}[h!]
    \begin{center}
        \includegraphics[width=0.95\linewidth]{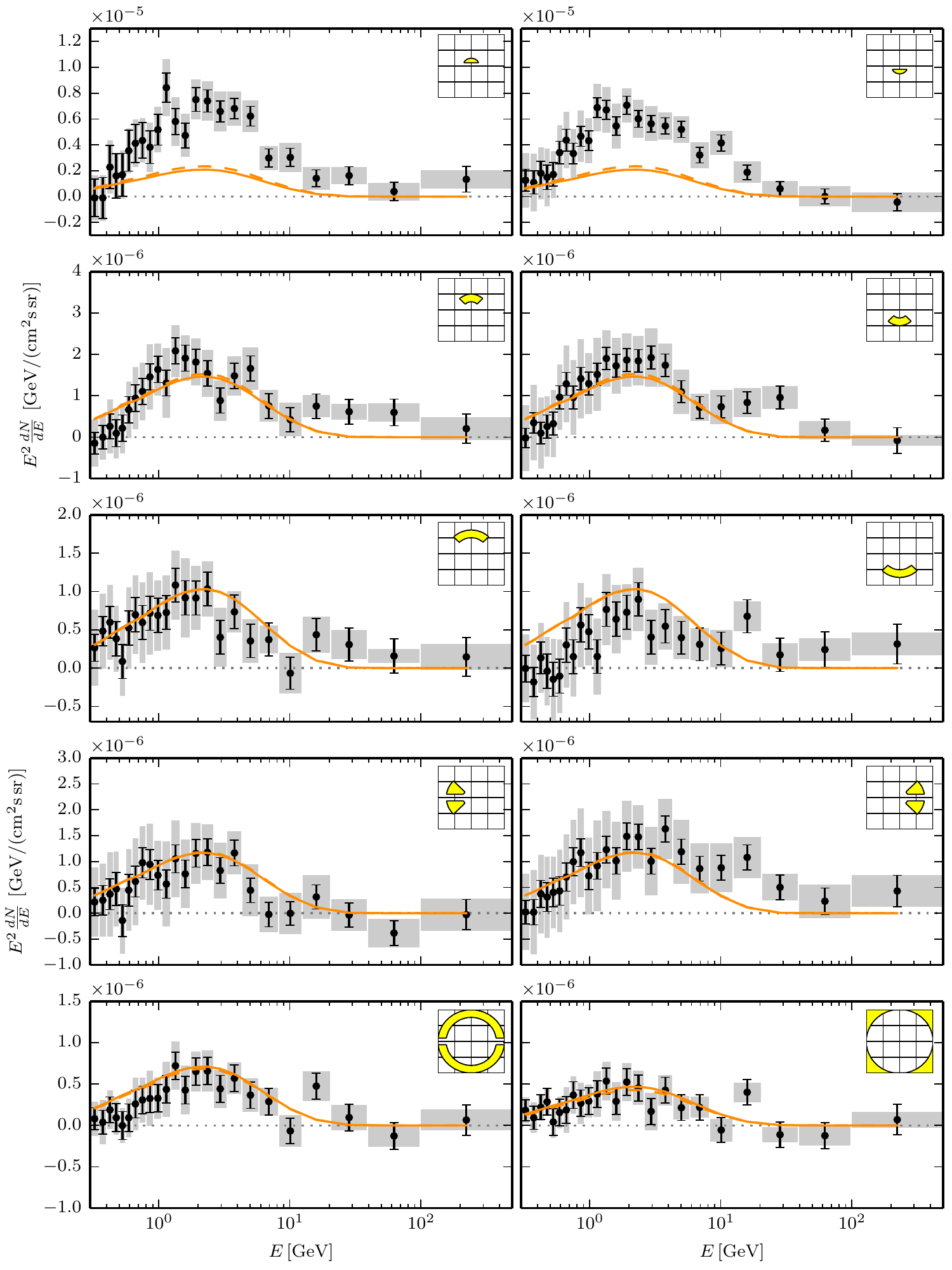}
    \end{center}
    \caption{Same as figure~\ref{fig:fluxes_HR} for the two MW-like selected haloes in the APOSTLE IR run.}
    \label{fig:fluxes_LGIR}
\end{figure}

As explained above, the DM density profile enters squared in the prediction of the
gamma-ray intensity flux and anisotropy signal that comes from annihilation of DM particles in the halo of the MW, 
see for example~\cite{Calore:2014hna}.
Searches for DM at the GC are thus especially sensitive to the uncertainty affecting
the determination of the inner distribution of the DM density. In this section, we
focus on the implications that the density profiles discussed above entail for the DM
interpretation of the \Fermi~GeV excess.

\medskip

We remind the reader that the analysis in~\cite{Calore:2014xka} tested several
templates for the GeV excess spatial distribution, i.e.~\emph{morphology}, assuming
an underlying generalised NFW (gNFW) profile:
\begin{equation}
\rho(r) = \frac{\rho_s}{(r/r_s)^{-\gamma} \, (1 + r/r_s)^{3-\gamma}} \, , 
\end{equation}
and varying the inner slope $\gamma$ in the range 0.7 -- 1.5 (thus accounting for
both shallower and steeper profiles at small radii than the traditional NFW).  The
best-fit value of $\gamma$ was found to be $1.26\pm0.15$, corresponding to an inner
slope steeper than the standard NFW profile.

\medskip

In order to verify the plausibility of this interpretation, it is crucial to test the
profiles predicted by hydrodynamic simulations against the GeV excess data.  In
particular, while we know already that the spectral shape of the signal is compatible
with a DM annihilation signal into light and b-quarks pairs (even when properly
taking into account the background model systematic
uncertainties)~\cite{Calore:2014nla}, the spatial distribution of the signal
represents the key-point-test for any model under scrutiny.  We use the GeV excess
data as derived by~\cite{Calore:2014xka} in the ten segments analysed in the region
$|l| < 2^{\circ}$ and $2^{\circ} < |b| <20^{\circ}$.  This region corresponds to a radial
scale from about 0.3 kpc up to 3 kpc.  Clearly, the lower limit is not probed by any
of our three simulations whose effective resolution, $2.8\times \epsilon$, is at
best 0.87 kpc. Therefore, to make predictions down to the scale at
which the GeV excess is measured we need to extrapolate the simulated profiles below
the resolution limit.  

Following a traditional approach to the extrapolation issue, we adopt a power-law
whose steepness is the maximal compatible with the total mass inside the extrapolation radius, 
namely the maximal asymptotic slope~\cite{Navarro:2004,Springel:2008cc}.
At each radius $r$,  mean and local densities define a robust upper limit on the asymptotic inner slope:
$\gamma_{\max}(r)= 3 (1 - \rho(r)/\bar{\rho}(r))$.
However, as already explained in section~\ref{sec:rhodm}, possible numerical effects still might occur
between the resolution limit and the Power radius. The Power radius was developed for the 
specific purpose of ascertaining the minimum radius at which the enclosed mass is converged~\cite{Power:2002sw}, and is
a useful guide for this test.
For this reason, we believe that only extrapolating 
from the Power radius guarantees the results to be stable against residual numerical effects.
Moreover, we checked that extrapolations from smaller radii (down the the effective resolution limit)
would lead to even shallower inner profiles.
Given our final aim to compare the simulated DM profiles with the GeV excess data (that require an inner slope as steep as 1.26), 
our choice is truly ``conservative" in a two-fold way: (a) the maximal asymptotic slope extrapolation guarantees that, at any radius $r$, 
profiles steeper than the maximal asymptotic slope are not allowed by the simulation data; 
(b) the choice of the Power radius guarantees that profiles steeper than the maximal asymptotic slope at the Power radius are
not allowed by the simulation data \emph{within the resolution/convergence limit of the simulation}.


The profile so determined is the
steepest power-law that can be accommodated by the simulation within its resolution/convergence limit.
 
The maximal asymptotic slope values in the simulation we analysed are: 
\begin{itemize}
\item EAGLE IR (10 haloes): $ 0.89 < \gamma_{\rm max} < 1.95$, at $R_{\rm P03}$= 3.6 kpc
\item  EAGLE HR (2 haloes): $0.94 < \gamma_{\rm max} < 0.98 $ at $R_{\rm P03}$= 1.8 kpc 
\item APOSTLE IR (2 haloes): $0.50 < \gamma_{\rm max} < 0.62$ at $R_{\rm P03}$= 1.8 kpc. 
\end{itemize}

Only among the relatively low-resolution EAGLE IR galaxies we find objects with a maximal slope as steep as the one required to explain the 
GeV excess emission, $1.26\pm0.15$, and that is only because the maximisation of the slope, if calculated from the relatively large Power
 radius of 3.6 kpc, leaves considerable freedom in the choice of the profile. 
For all other DM haloes (of EAGLE IR, EAGLE HR and APOSTLE IR selected MW-like galaxies) the maximal asymptotic
slopes at the Power radii are significantly shallower than the steep profile required to fit the data.
The result of our extrapolation indicates that no simulated halo has enough DM mass within 
the Power radius to support profiles as steep as $r^{-1.26}$ (as required by the data).
In the following discussion, we will consider \emph{only} high resolution haloes, namely EAGLE HR and
APOSTLE IR, since we believe that for those even a very conservative approach might lead to 
realistic results, contrary to what happens for EAGLE IR, where the resolution
  is too low to obtain useful constraints.

\medskip

We adopt the spherically averaged (and extrapolated) DM density
profiles of the 4 haloes in EAGLE HR and APOSTLE IR to perform a joint fit in the ten
regions analysed by~\cite{Calore:2014xka}. The aim is to assess how well the GeV excess
data are compatible with the spherically symmetric DM signal predicted by EAGLE simulations.
The DM signal is the one predicted by the DM density
distribution of the 4 selected MW-like galaxies, adopting the maximal asymptotic 
slope extrapolation. For a generic
WIMP DM candidate, the predicted flux of photons is written as:
\begin{equation}
\frac{d \Phi_{\gamma}}{dE} = 
  \frac{\langle \sigma
  v\rangle}{8\pi \,m_\chi^2} \, \frac{dN_{\gamma}}{dE}
  \int_\text{l.o.s.}\!\!\!\!\! ds\;\rho^2(r(s, \psi))\,.
  \label{eqn:fluxADM}
\end{equation}
The particle physics parameters ($\langle \sigma v \rangle$ and mass) are determined
by the fitting procedure (see below).  The integral along the line of sight (l.o.s.) 
of the DM density profile squared (which depends on the l.o.s.~variable $s$ and on
the angle from the GC, $\psi$), the so-called J-value, instead is fixed for each
selected galaxy.  We fix the branching ratio to be 100\% annihilation into b-quarks
(our conclusions are however independent of the assumed annihilation channel) and
we compute the corresponding gamma-ray spectrum, $dN_{\gamma}/dE$, with
\texttt{DarkSUSY 5.1.1}~\cite{Gondolo:2004sc}.  The fitting procedure fully takes
into account the background model systematic uncertainties in each of the ten
sub-regions, but it does not account for possible correlations between different
sub-regions.  The inclusion of those correlated uncertainties has been demonstrated to allow
more freedom for models fitting the excess, as thoroughly shown in the case of DM
in ref.~\cite{Calore:2014nla}.  The $\chi^2$ function is:
\begin{equation}
\label{eq:chi2}
    \chi^2 = \sum_{i=1}^{10} \sum_{j,k=1}^{24} 
    (d_{ij} - \mu_{ij} ) (\Sigma_{jk}^{i})^{-1}
    (d_{ik} - \mu_{ik} )\;,
\end{equation}
where $d_{ij}$ ($\mu_{ij}$) represents the measured (predicted) flux in the segmented
region $i$ and energetic bin $j$; $\Sigma_{jk}^i$ is the covariance matrix for energy
bins $j$ and $k$ in region $i$~\cite{Calore:2014xka}.  The $p$-values quoted below
refer to this $\chi^2$ function for the ten sub-regions fit, assumed to follow a
$\chi^2_k$ distribution with $k=240-2$ degrees of freedom.

The free parameters of the fit, $\langle \sigma v \rangle$ and DM mass, are determined 
by minimising the $\chi^2$ in eq.~\ref{eq:chi2}. We emphasise that any difference in the absolute normalisation 
of the DM density profile is re-absorbed in the fitting procedure by a different combination of annihilation cross-section
and DM mass that still provides a good fit to the data.
The goodness of fit is only determined by the shape of the profile.

In table~\ref{tab:svm}, we quote the best-fit values for $\langle \sigma v \rangle$
and mass, as well as the $p$-values, for the best-fit galaxies in the EAGLE HR and APOSTLE IR run
(and for a gNFW profile with $\gamma$=1.26).  Correspondingly, we show constraints in
the ($\langle\sigma v\rangle$, $m_\chi$) plane from the fit to the ten sub-regions in
figure~\ref{fig:contours}.  As expected, owing to the lower J-value due to the
flattening in the inner kiloparsecs, the best-fit cross section in the case of a DM
density profile drawn from the EAGLE HR and APOSTLE IR selected MW-like galaxies is slightly higher
than the best-fit cross section obtained when adopting a gNFW ($\gamma=1.26$) in the
fit.\footnote{We note that an additional source of discrepancy in the overall normalisation 
of the contour regions in figure~\ref{fig:contours} is due to the different local dark matter density values.}
Although it has been shown that current upper limits on the annihilation cross
section from dwarf spheroidal galaxies are not yet in tension with the DM parameter
space preferred by the DM interpretation of the GeV excess when a gNFW profile with
$\gamma=1.26$ is assumed~\cite{Calore:2014nla}, we note that the results obtained by
using a shallower DM profile might lead sooner to a tension with dwarf limits.

Figures~\ref{fig:fluxes_HR} and \ref{fig:fluxes_LGIR} show the fluxes in the ten
sub-regions for the density profiles of the 4 MW-like galaxies and particle physics
parameters as obtained by the fit for each galaxy.  In general, all DM haloes of selected galaxies
fail in accounting for the flux in the two innermost regions, that extend up to $5^{\circ}$
(0.75 kpc) above and below the Galactic plane. This is due to the flattening of the
profile on scales of 1 -- 2 kiloparsecs, even when extrapolating the density profile
with the maximal asymptotic slope method.  On the other hand, in the outermost
regions the fit is relatively good because the profile preferred
by the data and that used by the model are similar.

\begin{figure}
    \begin{center}
        \includegraphics[width=0.6\linewidth]{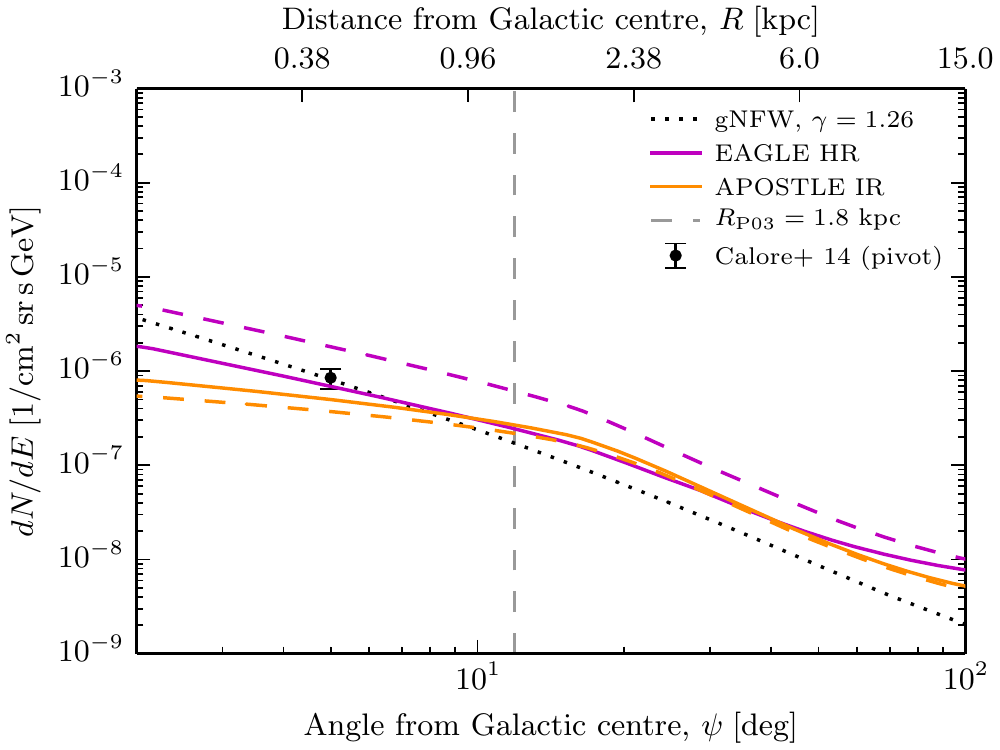}
    \end{center}
    \caption{The \emph{magenta} (\emph{orange}) lines represent the
      angular profile of the DM signal predicted by the selected MW-like halo's
      density profiles in the EAGLE HR (APOSTLE IR) run. The 
      \emph{solid} line refers to the best-fit halo, while the \emph{dashed} line
      corresponds to the density profile of the other halo selected normalised 
      to the best-fit cross section and mass of the best-fit halo. The difference between  
      two lines of the same colour thus indicates the 
      uncertainty affecting the J-value. The \emph{black dotted line} represents the expectation for the gNFW
      profile with $\gamma$=1.26 -- that best fits the GeV excess data --, 
      while the \emph{black point} is the flux of the GeV
      excess at $5^{\circ}$ away from the GC (that was shown to be a good pivot point
      in~\cite{Calore:2014xka}). The \emph{vertical grey dashed} line represents the
      Power radius for the two simulation runs.}
    \label{fig:angular_IR_HR}
\end{figure}

In figure~\ref{fig:angular_IR_HR} we display the angular profile of the DM
signal as predicted when using the density profile of the best-fit MW-like galaxies in
the EAGLE HR and APOSTLE IR runs. The 
set of 4 lines refers to the 4 galaxies. For galaxies in the same simulation run the
particle physics factor is fixed to the best-fit parameters of the best-fit galaxy of that simulation.
The scatter between two lines of the same colour thus indicates the 
uncertainty affecting the J-value, including the difference in the local DM density.  
For comparison, the
expected DM signal for the gNFW profile with $\gamma$=1.26 is also shown.  The data
point represents the flux of the GeV excess at 5$^{\circ}$ from the GC. It was
demonstrated in~\cite{Calore:2014xka} that this is a good pivot point since the flux
only mildly depends on the inner slope of the DM profile.  Below 5$^{\circ}$
the expected signal from EAGLE HR and APOSTLE IR selected MW-like galaxies does not account for the
extra-emission at the GC.  The mild discrepancy between all simulated haloes' angular profiles and the
gNFW prediction for $R > 2.4$ kpc is due to the
different local density normalisation. 

\medskip

\begin{figure}
    \begin{center}
        \includegraphics[width=0.5\linewidth]{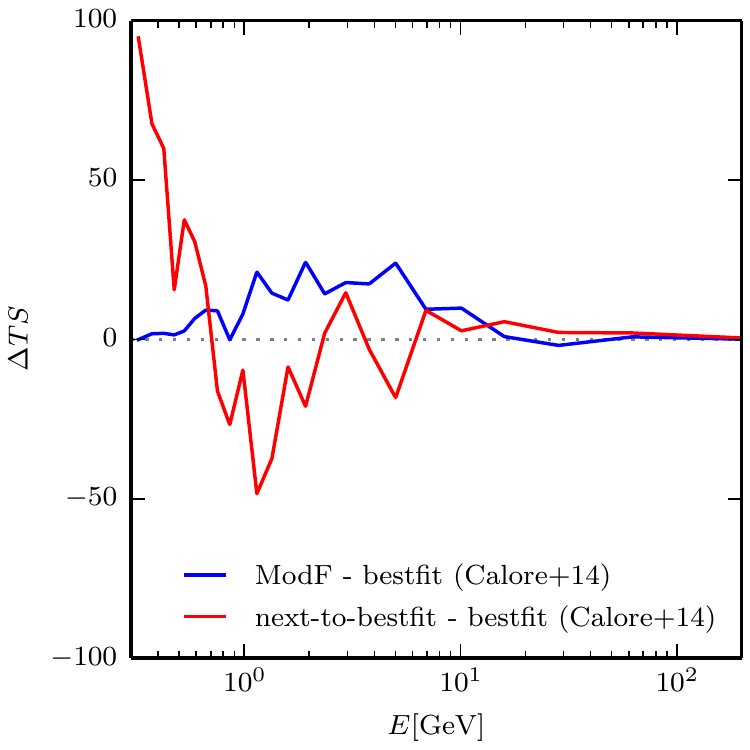}
    \end{center}
    \caption{Variation of the TS($= -2 \ln \mathcal{L}$) between the best-fit Galactic diffuse emission model
      for the EAGLE HR DM profile template and the best-fit Galactic diffuse emission model
      for the gNFW template~\cite{Calore:2014xka}: the \emph{blue solid} line 
      uses as DM template the EAGLE HR density
      profile extrapolated with the maximal asymptotic slope method from the Power radius down to smaller
      scales. The TS
      variation of the next-to-best-fit Galactic diffuse emission model for the gNFW
      template with respect to the best-fit model is shown by the \emph{red line}.}
    \label{fig:TS_HR}
\end{figure}

Given the DM density profile predicted by the hydrodynamic simulation (and the fact
that those profiles are not well described by any of the parameterisations commonly
adopted in the literature), we re-analyse \Fermi-LAT gamma-ray data using a GeV excess template built
on the spatial distribution of the galaxy that best fits the GeV excess data, i.e.~the
best-fit galaxy of the EAGLE HR run.  In particular, we are interested in a
quantitative estimate of the goodness of fit of such a DM template with respect to
the best-fit template adopted in~\cite{Calore:2014xka}.  As thoroughly explained
in~\cite{Calore:2014xka}, such an analysis depends on the Galactic diffuse model
adopted.  We run the fitting template procedure described in~\cite{Calore:2014xka}
for the DM template based on the EAGLE HR DM halo profile \emph{extrapolated} with
maximal asymptotic slope at the Power radius of 0.94.  We analyse all 60 models for
the Galactic diffuse emission presented in~\cite{Calore:2014xka}.  We find that the
Galactic diffuse emission model leading to the best fit for the EAGLE HR DM halo template
is the so-called model F, the same foreground model giving the best-fit result for a
gNFW template with $\gamma=1.26$.  In figure~\ref{fig:TS_HR}, we show the variation
of the test statistic\footnote{The TS is 
  a measure of the goodness of fit, being directly related to the likelihood.} TS($=
-2 \ln \mathcal{L}$), $\Delta$TS, as function of the energy for the best-fit Galactic
diffuse model for the EAGLE HR DM halo template when compared to the best-fit one for the
gNFW template with $\gamma=1.26$.  The $\Delta$TS is an indicator of how much better
a model performs with respect to another at all energies above 1 GeV.  It is evident
that in general the EAGLE HR DM template leads to a worse fit in the energy range
relevant for the GeV excess (1 -- 3 GeV), while it performs similarly to the gNFW
template at higher energy.  To give a term of comparison for the $\Delta$TS values,
we show in the same figure the $\Delta$TS between the next-to-best-fit and the
best-fit Galactic diffuse emission model when using the gNFW template with
$\gamma=1.26$. The variation in $\Delta$TS due to the different Galactic diffuse
emission model are as large as 50, although the differences are more pronounced at
low energies.  Remarkably, the uncertainties due to the variation of the DM
template are comparable to the ones due to the Galactic diffuse modelling.

\section{Conclusions}
\label{sec:conclusion}
In this work, we have used the set of cosmological, hydrodynamic simulations of the
EAGLE~\cite{Schaye:2015} and APOSTLE~\cite{2014arXiv1412.2748S} projects.  
We extracted the simulated
haloes in the generous Milky Way-like mass range of
$5\times10^{11}<M_{200}/\Msun<1\times10^{14}$.  For each resolution run, we required
that the galaxies satisfy observational properties of the Milky Way, besides the
uncertain constraint on the halo mass.  In particular, we defined galaxies to be
\emph{Milky Way-like} if (i) they give a good fit to the Galactic rotation
curve~\cite{Iocco:2015xga}, (ii) have a stellar mass in the 3$\sigma$ observed
Milky Way stellar mass range~\cite{McMillan:2011wd}, and (iii) show a dominant disc
in the stellar component.  We then analysed the dark matter density profiles of the
selected Milky Way-like galaxies and we derived implications for the dark matter
interpretation of the \Fermi~GeV excess.  We summarise our findings below:

\begin{itemize}

\item[(i)] The adopted selection criteria proved to be very powerful in reducing the
  large variation in the rotation curves predicted by simulated galaxies selected on the
  basis of the virial mass only (\cf section~\ref{sec:selection}).

\item[(ii)] As a consequence, the dark matter density profiles of the Milky Way-like
  galaxies in our final selection are remarkably consistent with each other, and they
  have local dark matter densities easily compatible with the measured value of
  about 0.4 GeV/cm$^3$.

\item[(iii)] The subgrid physics model implemented in EAGLE (and APOSTLE) simulations leads to
  the dark matter density profiles of Milky Way-like galaxies exhibit a flattening 
  in the inner 1.5 -- 2 kpc (probably due to starburst events at low
  redshift) and a regime of contraction, from 1.5 -- 2 kpc up to about 10
  kpc, where the profile's slope is steeper than 1. While the flattening in the EAGLE IR simulation might still be affected by numerical effects, 
  the flattening seen in EAGLE HR and APOSTLE IR appears to have a physical origin,
  \cf section~\ref{sec:rhodm}.

\item[(iv)] We use the dark matter density profiles of the selected Milky Way-like galaxies to compute
  the predicted gamma-ray flux from dark matter annihilation. We extrapolate the dark matter profiles
  down to 0.3 kpc through the maximal asymptotic slope method.
  Even under the very conservative assumption of extrapolating from the Power radius, 
  when performing the fit
  to the GeV excess data~\cite{Calore:2014xka}, we found that those dark matter profiles fail to
  reproduce the right morphology of the excess in the innermost regions (within
  $5^{\circ}$ above and below the Galactic plane).\footnote{However, we warn the reader
  that, at the moment, we cannot really rule out that the shallower profile towards the centre 
  is due to numerical or subgrid physics effects~\cite{Schaye:2015}.}
  On the other hand, all  
  selected Milky Way-like galaxies give a good fit to the GeV excess data in the other regions of interest. 
  Moreover, the parameter
  space for annihilation cross section and dark matter mass are well in agreement
  with previous findings within 20\%.

\end{itemize}

We thus showed that the dark matter density profiles of our selected Milky Way-like galaxies lead to a gamma-ray flux
from dark matter annihilation in the main Galactic halo that cannot
fully account for the observed \Fermi~GeV excess. Our results do not exclude the possibility that dark 
matter annihilations contribute to the GeV excess, but if that is the case, they require an additional component of emission in the innermost few degrees,
e.g. in the form of point-like sources whose fluxes are too dim to be detected by the \Fermi~LAT telescope as individual
objects, as recently proposed in Refs.~\cite{Bartels:2015aea,Lee:2015fea}.

\paragraph*{Acknowledgements.}
We thank Arianna Di Cintio and Cristoph Weniger for useful discussions.
Special thanks to Fabio Iocco and Miguel Pato 
for providing the extensive compilation of rotation curve 
measurements used in this paper and for many fruitful discussions.
G.B. (P.I.), N.B.~and F.C.~acknowledge support from the European Research Council
through the ERC starting grant WIMPs Kairos. This work used the DiRAC Data Centric
system at Durham University, operated by the Institute for Computational Cosmology on
behalf of the STFC DiRAC HPC Facility (www.dirac.ac.uk). This equipment was funded by
BIS National E-infrastructure capital grant ST/K00042X/1, STFC capital grant
ST/H008519/1, and STFC DiRAC Operations grant ST/K003267/1 and Durham
University. DiRAC is part of the National E-Infrastructure. This work is part of the
D-ITP consortium, a program of the Netherlands Organisation for Scientific Research
(NWO) that is funded by the Dutch Ministry of Education, Culture and Science (OCW).
This work was supported by the Science and Technology
Facilities Council (grant number ST/F001166/1);
European Research Council (grant numbers GA 267291 
``Cosmiway'' and GA 278594 ``GasAroundGalaxies'') and by 
the Interuniversity Attraction Poles Programme initiated by the
Belgian Science Policy Office (AP P7/08 CHARM).
R.A.C.~is a Royal Society Research Fellow.

\clearpage
\appendix
\section{Variation with halo mass}
\label{app:M200}
\begin{figure}
    \begin{center}
        \includegraphics[width=0.48\linewidth]{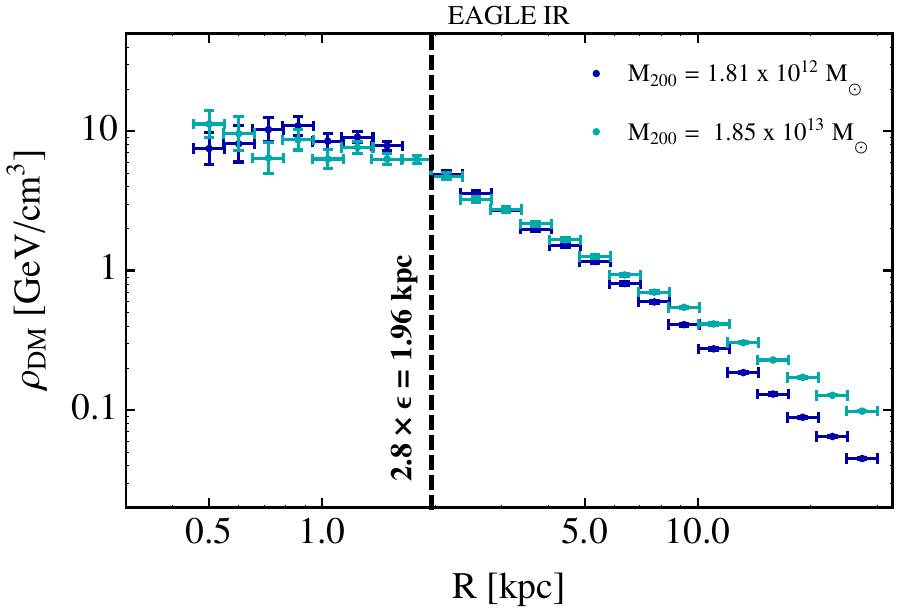}
        \includegraphics[width=0.48\linewidth]{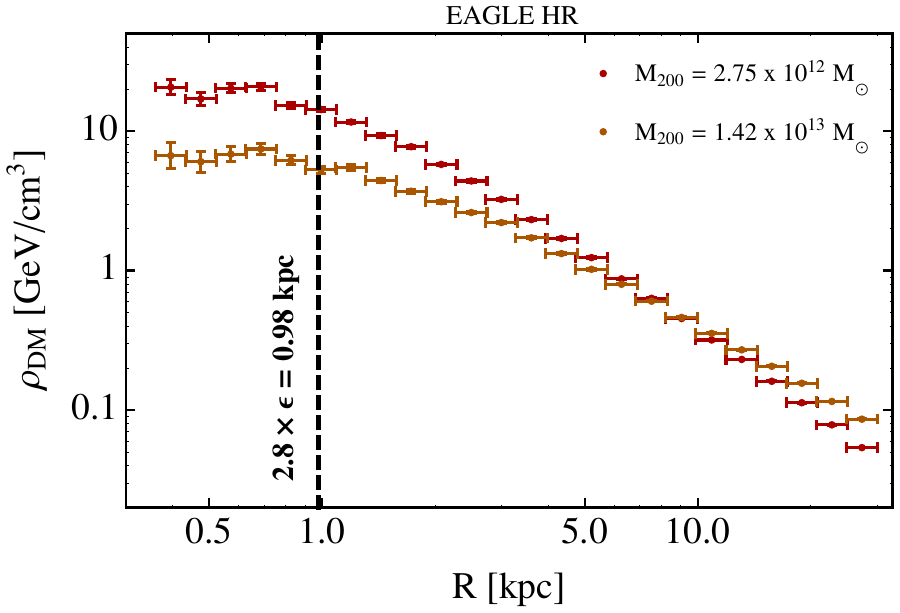}
    \end{center}
    \caption{DM density profiles for the two selected haloes with the correct total
      stellar mass (within 3$\sigma$) and with the lowest and largest halo mass,
      $M_{200}$, for the EAGLE IR (\emph{left} panel) and EAGLE HR (\emph{right} panel) runs.}
    \label{fig:M200rho}
\end{figure}

In figure~\ref{fig:M200rho}, we display the DM density profile for two haloes in the
EAGLE IR and EAGLE HR runs that have the correct MW observed stellar mass range (within
3$\sigma$) and that have the lowest and highest $M_{200}$ respectively. This figure
illustrates that the features of the DM density profile (i.e.~the inner flattening and the contraction regime) 
do not strongly depend on $M_{200}$. 
While for EAGLE IR also the normalisation of the density profiles is well 
matched, for EAGLE HR haloes the overall profile normalisation might differ up to a factor of 5 at the 
resolution limit. 
Nevertheless, for the purposes of our analysis, such a normalisation discrepancy is not critical.
Indeed, as explained in section.~\ref{sec:idm}, the goodness of fit to the GeV excess data is driven by the
shape of the DM density profile in the range 0.3 -- 3 kpc and by the extrapolated maximal asymptotic slope.
To demonstrate that the halo mass variation does not introduce any bias in our analysis, we perform the
fit to the GeV excess data for the halo with the largest mass in figure~\ref{fig:M200rho} (right panel).
The best-fit parameters are $\langle \sigma v \rangle =  (2.45 \pm 0.20) \times 10^{-26} \, \rm cm^{3}/s$ 
and $m_{\chi}= 45.4 \pm 2.1 \, \rm GeV$. Those values are compatible (within 2$\sigma$) with the EAGLE HR entry in
table~\ref{tab:svm}.
In light of this result, we are confident that discrepancies in the halo mass does not alter our final conclusion.

\section{Varying local Galactic parameters}
\label{app:v0R0}

Throughout this paper, we have assumed $R_0=8$ kpc, $v_0=230$ km/s, and the peculiar
velocity of the Sun, $(U, V, W)_\odot=(11.10, 12.24, 7.25)$ km/s given in Galactic
coordinates. In this appendix, we consider different choices of $R_0$, $v_0$ and
$V_\odot$, showing that the main conclusions of the paper remain unchanged. We follow
ref.~\cite{Iocco:2015iia}, and consider the following four different configurations:

\vspace{10pt}
\noindent
(a) $R_0=7.98$ kpc, $v_0=214.44$ km/s for $V_\odot=26$ km/s;\\
(b) $R_0=7.98$ kpc, $v_0=236.94$ km/s for $V_\odot=5.25$ km/s;\\
(c) $R_0=8.68$ kpc, $v_0=235.53$ km/s for $V_\odot=26$ km/s;\\
(d) $R_0=8.68$ kpc, $v_0=258.19$ km/s for $V_\odot=5.25$ km/s.\\

\begin{figure}
    \begin{center}
        \includegraphics[width=0.48\linewidth]{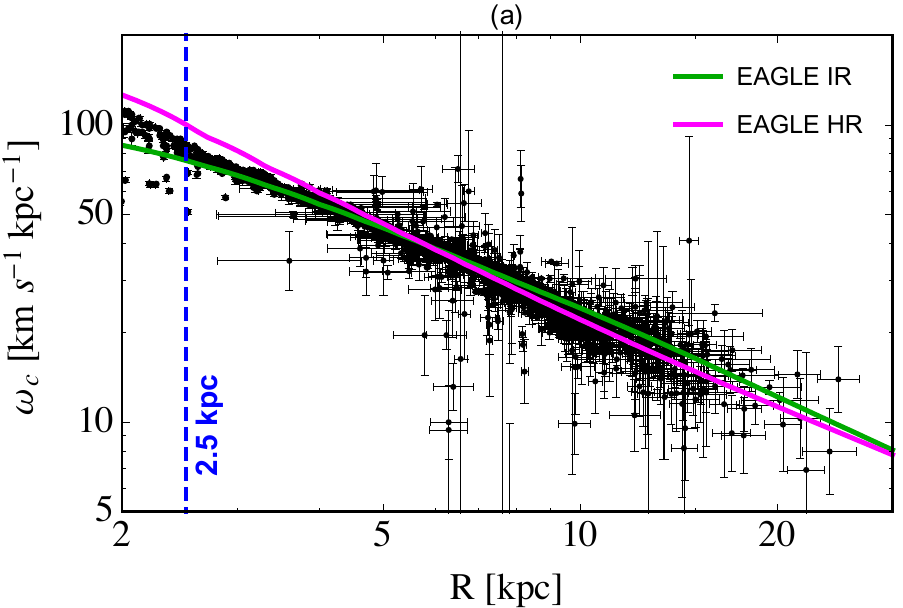}
        \includegraphics[width=0.48\linewidth]{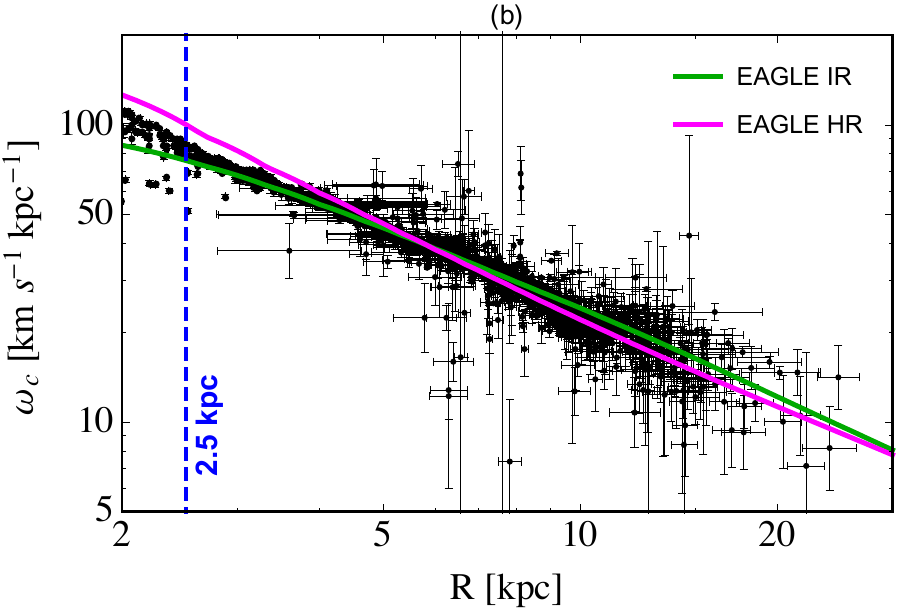}\\
         \includegraphics[width=0.48\linewidth]{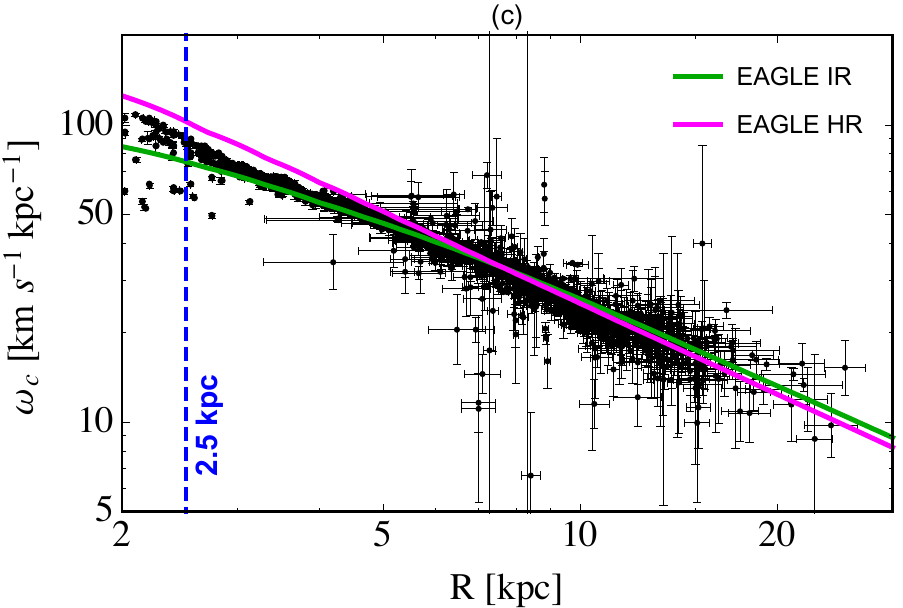}
        \includegraphics[width=0.48\linewidth]{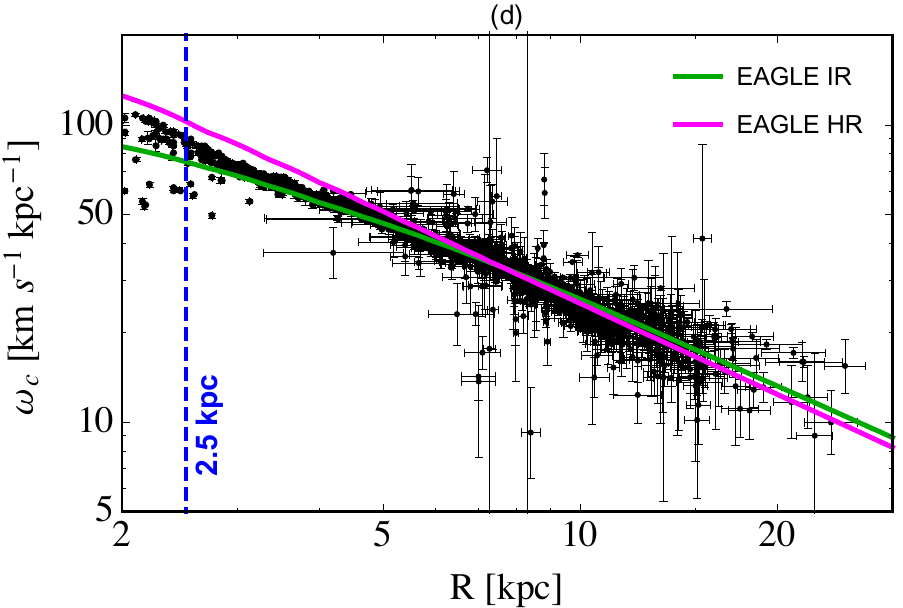}
    \end{center}
    \caption{Observed MW angular circular velocity as a function of galactocentric
      distance from~\cite{Iocco:2015xga} (\emph{black} points and error bars) for
      configuration (a) (\emph{top left}), (b) (\emph{top right}), (c) (\emph{bottom
        left}), and (d) (\emph{bottom right}), as well as the best-fit halo in the EAGLE HR
      (\emph{magenta}) and EAGLE IR (\emph{green}) runs.}
    \label{fig:BestFit-R0v0}
\end{figure}

\begin{figure}
    \begin{center}
        \includegraphics[width=0.48\linewidth]{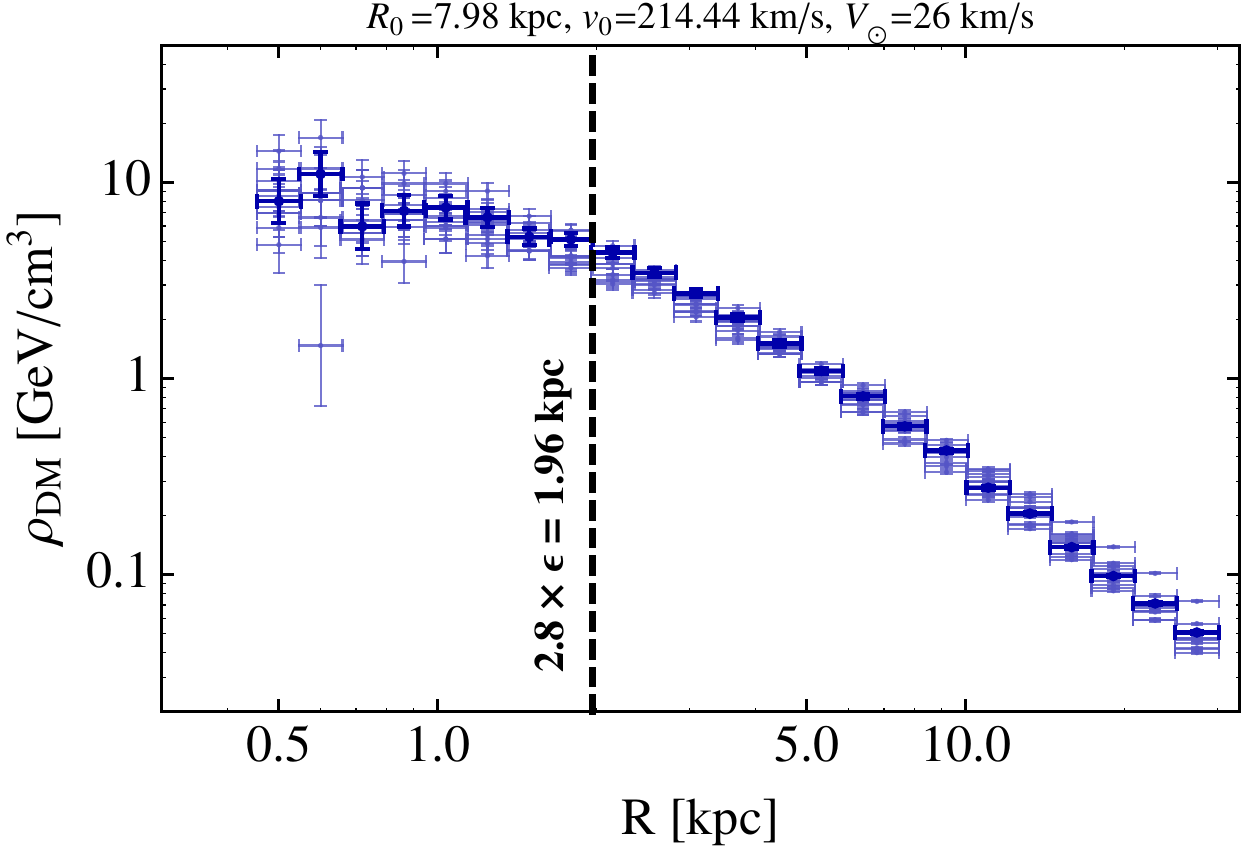}
        \includegraphics[width=0.48\linewidth]{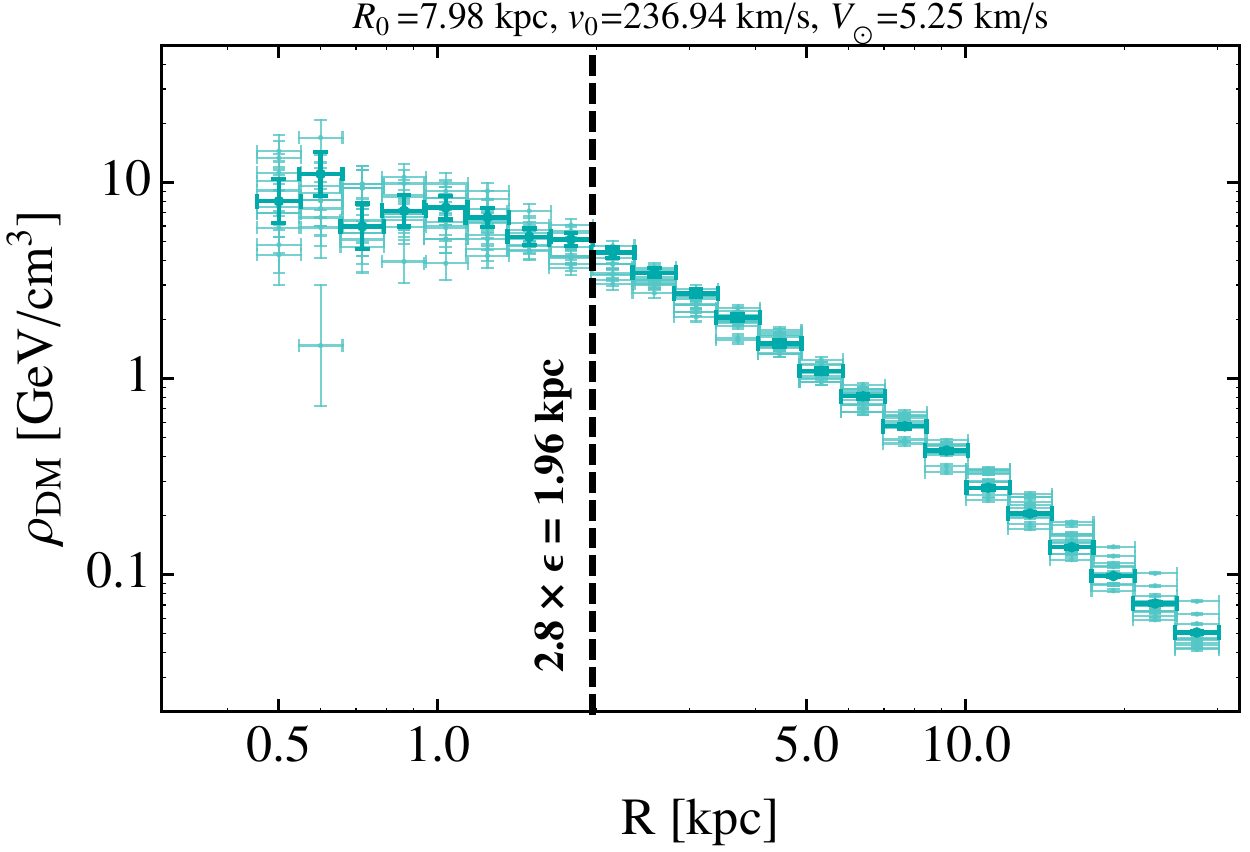}\\
         \includegraphics[width=0.48\linewidth]{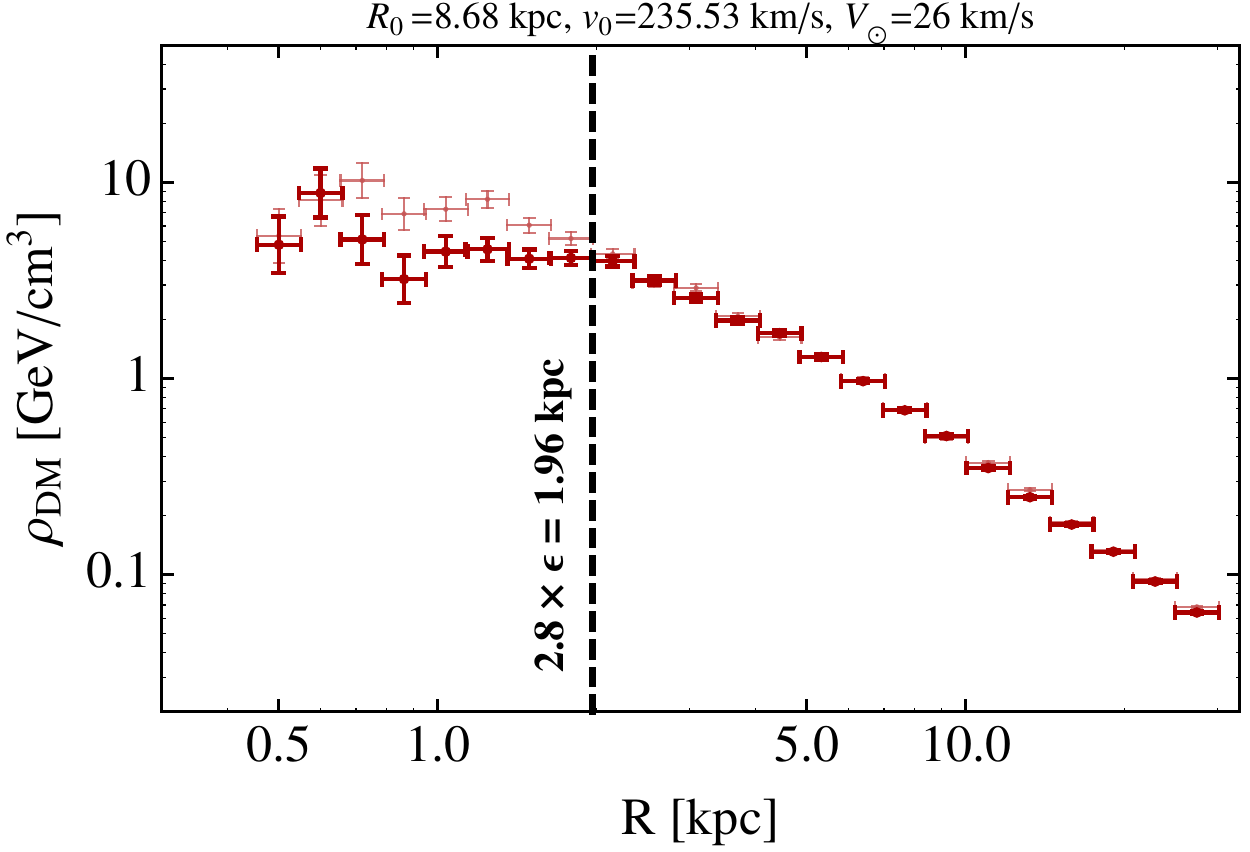}
        \includegraphics[width=0.48\linewidth]{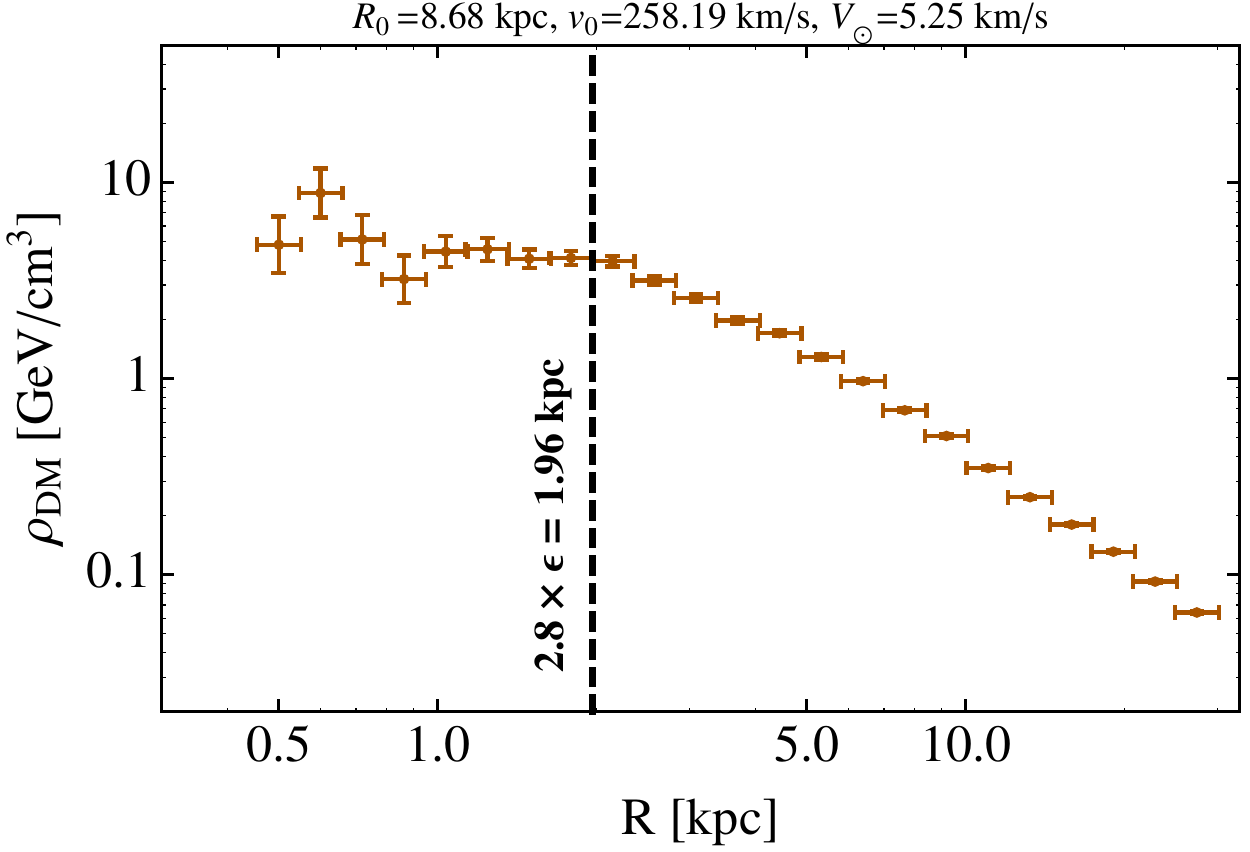}
    \end{center}
    \caption{Same as figure~\ref{fig:profile_IR_HR} for the EAGLE IR run and for the four combinations of $v_0$ and $R_0$.}
    \label{fig:BestFit-R0v0-rho}
\end{figure}

\begin{figure}
    \begin{center}
        \includegraphics[width=0.48\linewidth]{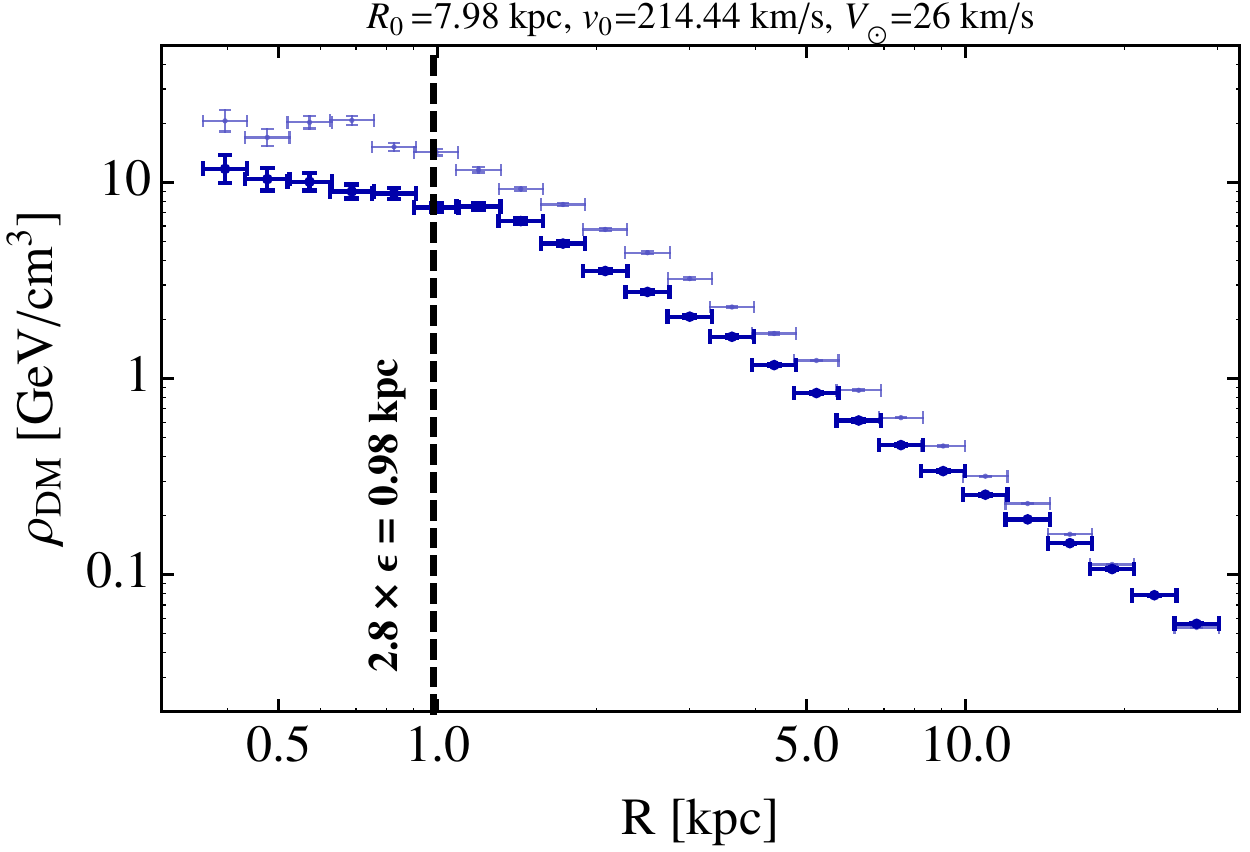}
        \includegraphics[width=0.48\linewidth]{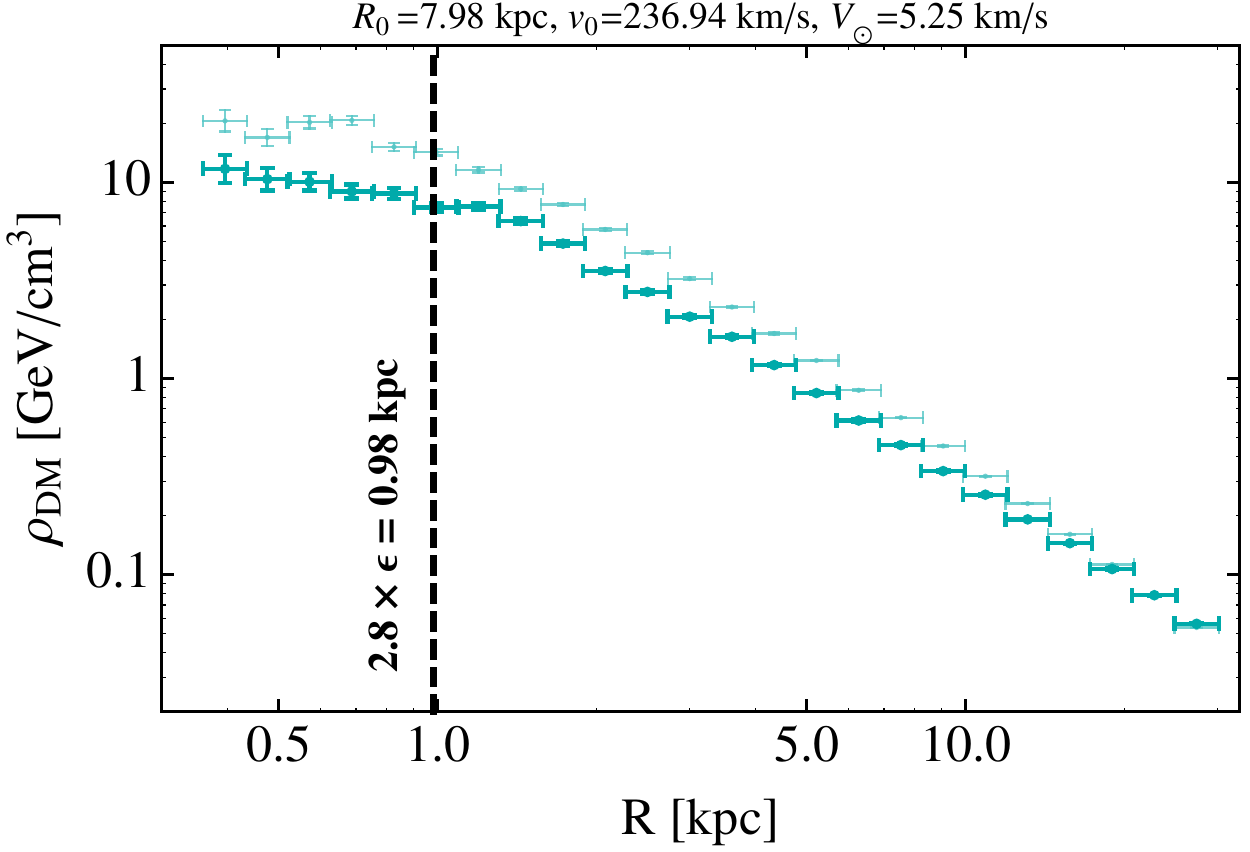}\\
         \includegraphics[width=0.48\linewidth]{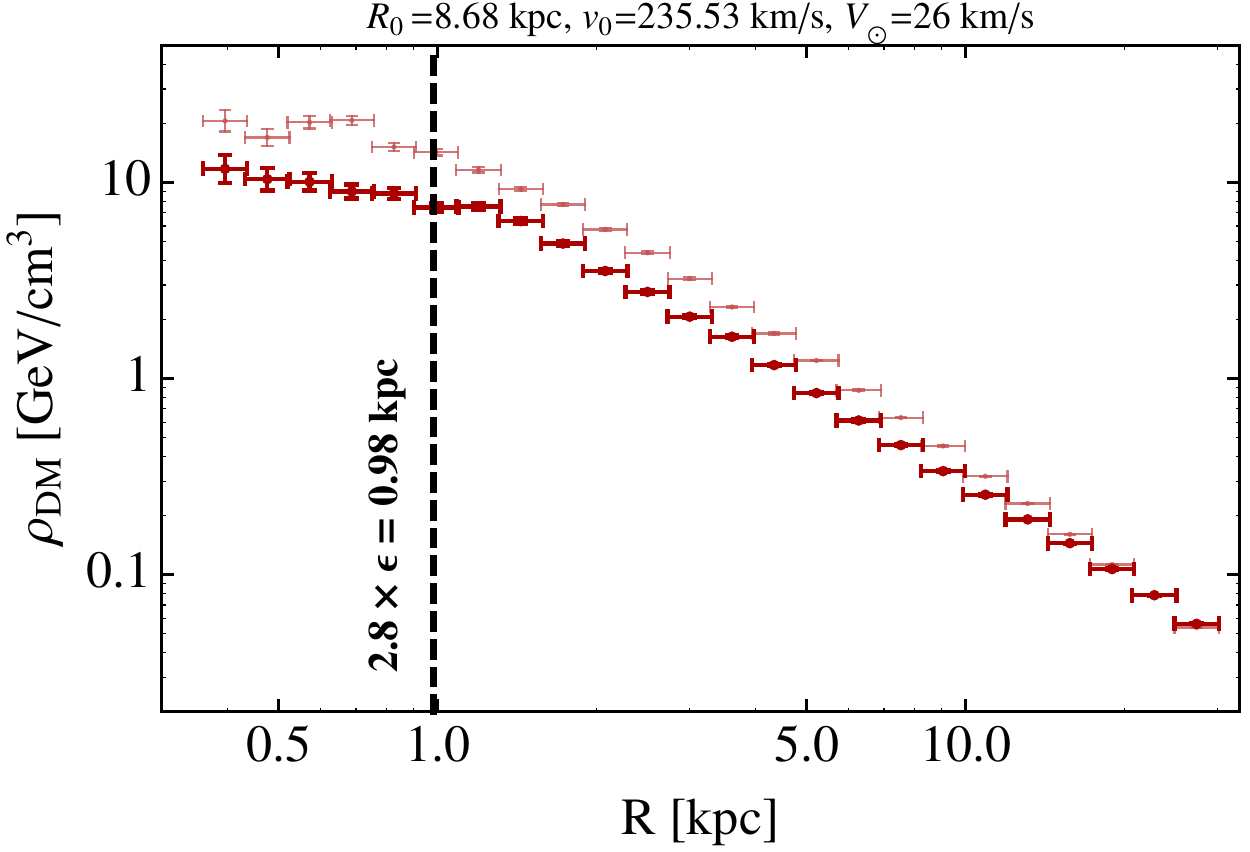}
        \includegraphics[width=0.48\linewidth]{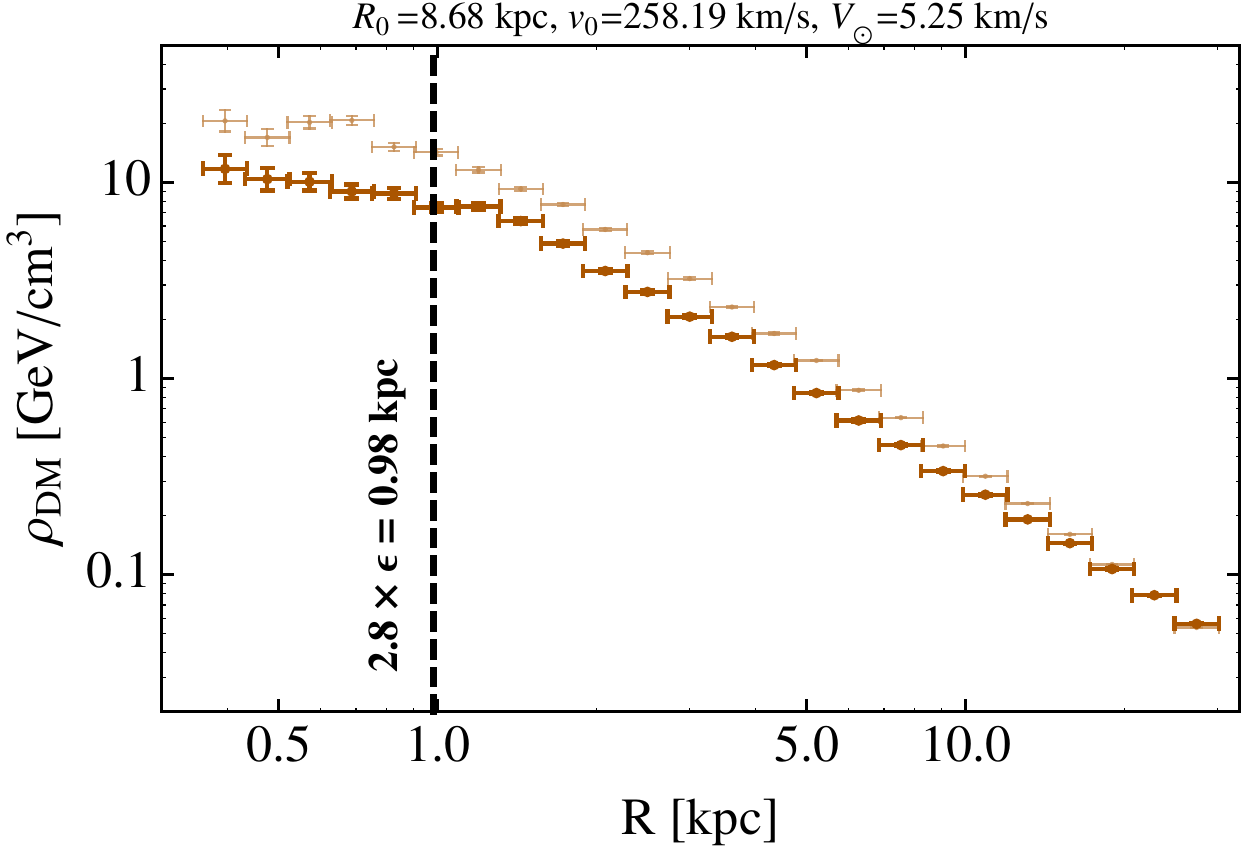}
    \end{center}
    \caption{Same as figure~\ref{fig:profile_IR_HR} for the EAGLE HR run and for the four combinations of $v_0$ and $R_0$.}
    \label{fig:BestFit-R0v0-HR}
\end{figure}

These configurations are used to obtain the observed MW rotation curve data. We
perform the $\chi^2$ analysis discussed in section~\ref{sec:rotcurves} to find the
best-fit haloes which satisfy the criteria in section~\ref{sec:selection}. The
observational data for the MW angular circular velocity as well as the best-fit
haloes for the EAGLE HR and EAGLE IR runs are shown in figure~\ref{fig:BestFit-R0v0}. The four
panels of the figure correspond to the four configurations. The reduced $\chi^2$ for
these haloes are specified in table~\ref{tab:R0v0}.  The DM density profiles of the
MW-like selected haloes are shown in figures~\ref{fig:BestFit-R0v0-rho} and
\ref{fig:BestFit-R0v0-HR} for the EAGLE IR and EAGLE HR runs, respectively.  All haloes (both in
the EAGLE IR and EAGLE HR runs) show a significant deviation from the NFW profile and the common
features of the density profile discussed in section~\ref{sec:rhodm} can be found
also when varying $R_0$, $v_0$ and $V_\odot$ parameters.

\begin{table}
    \centering
    \begin{tabular}{|c|c|c|c|}
      \hline
       Configuration & $\chi^2/(N-1)$, EAGLE HR & $\chi^2/(N-1)$, EAGLE IR \\
      \hline
  (a)      & 70.48 & 15.88 \\
       \hline
  (b)      & 73.21 & 16.15 \\
     \hline
   (c)      & 98.49  & 16.83 \\
    \hline
    (d)      & 93.62  & 17.92\\
      \hline
    \end{tabular}
    \caption{Reduced $\chi^2$ values for the best fit haloes which satisfy the criteria in section~\ref{sec:selection} in the EAGLE HR and EAGLE IR runs
      for the four configurations discussed in the text.}
    \label{tab:R0v0}
  \end{table}

\clearpage
\bibliographystyle{JHEP}
\bibliography{biblio}

\end{document}